\documentclass[draft]{agujournal2019}
\usepackage{url} 
\usepackage{lineno}
\usepackage[inline]{trackchanges} 
\usepackage{soul}
\usepackage{amsmath}
\usepackage{amsfonts}
\usepackage{amssymb}
\usepackage{graphicx}
\usepackage{verbatim}

\draftfalse

\journalname{JGR: Space Physics}

\begin{document}

\title{Effect of Electron Precipitation on E-Region Instabilities: Theoretical Analysis}

\authors{Y. S. Dimant\affil{1}, G. V. Khazanov\affil{2}, and M. M. Oppenheim\affil{1}}

\affiliation{1}{Center for Space Physics, Boston University, Boston, MA USA}
\affiliation{2}{NASA Goddard Space Flight Center, Greenbelt, MD USA}




\correspondingauthor{Yakov Dimant}{dimant@bu.edu}

\begin{keypoints}
\item During geomagnetic storms, strong electric fields and intense electron precipitation may overlap in the E-region ionosphere, where magnetospheric currents close.
\item Without precipitation, sufficiently strong electric fields drive E-region instabilities, leading to plasma turbulence and increased ionospheric conductance.
\item 
The intense electron precipitation may raise dramatically the instability threshold, largely suppressing the instability inside the auroral regions.
\end{keypoints}

\begin{abstract}
    During periods of strong geomagnetic activity, intense currents flow  from the magnetosphere into the high-latitude E-region ionosphere along geomagnetic field lines, B. In this region, collisions between the plasma and neutral molecules allow currents to flow across B, enabling the entire magnetosphere-ionosphere current system to close. These same currents cause strong DC electric fields in the E-region ionosphere where they drive plasma instabilities, including the Farley-Buneman instability (FBI). These instabilities give rise to small-scale plasma turbulence that modifies the large-scale ionospheric conductance that, in turn, affects the evolution of the entire near-Earth plasma environment. Also, during geomagnetic storms, precipitating electrons of high energies, $\gtrsim$~5 keV, frequently penetrate down to the same regions where intense currents and E fields develop. This research examines the effects of precipitating electrons on the generation of the FBI and shows that, under many common conditions, it can easily suppress the FBI in a predictable manner. Applying the kinetic SuperThemal Electron Transport (STET) code, we have analyzed the distribution function expected in the E-region and the effects of these energetic electron distributions on the FBI onset criterion. This shows that the plasma pressure of superthermal electrons may be comparable to, or even significantly exceed, the regular plasma pressure of the cold ionospheric plasma. This will increase the FBI threshold and suppress the instability in auroral regions. However, our detailed theoretical analysis shows that the effect of the superthermal precipitating electrons is much stronger than just the effect of the additional pressure. We were surprised to discover that the energy dependence of the electron-neutral collision frequency can greatly enhance the effect of this additional pressure, further suppressing the FBI, even at a moderate precipitation level. Therefore, we expect precipitation to exert a significant feedback on the magnetosphere by preventing the elevated conductivity caused by FBI driven turbulence. This suppression should be taken into account in global modeling of the magnetosphere-ionosphere coupling.
\end{abstract}


%
%

%


%
%
%
%

\section{Introduction\label{Introduction}}
During periods of intense geomagnetic activity, strong DC electric fields,
$\vec{E}_{0}$, perpendicular to the geomagnetic field, $\vec{B}$, penetrate
from the Earth's magnetosphere into the high-latitude ionosphere where they
dissipate energy, form electrojets, and drive plasma instabilities in the
E-region ionosphere, at altitudes roughly between 90 and 120~km. In the global
picture of magnetosphere-ionosphere coupling, this is the region where
most of the field-aligned magnetospheric currents close. The E-region
instabilities generate plasma density irregularities, typically within the
wavelength range from tens of centimeters to tens of meters, coupled with
wavelike electrostatic field fluctuations. The density irregularities have
been routinely detected as strong coherent radar echoes
\cite<e.g.,>{%
Bahcivan:30MHz05,Bahcivan:Observations06,Hysell:Imaging08,Forsythe:Dual}.
The electrostatic field fluctuations have been
detected by rocket flights through the lower ionosphere
\cite{Pfaff:Electric84,Pfaff:Electric87b,Pfaff:E92,Pfaff:Guara97,Rose:ROSE92,Fukao:SEEK98}.
The E-region instabilities include the Farley-Buneman
\cite{Farley:Ejet63,Buneman:Ejet63}, gradient
drift \cite{Hoh:63,Maeda:Theoretical63}, and
thermal instabilities \cite<e.g.,>[and references therein]{Dimant:Physical97,Dimant:Ion04,Oppenheim:Newly20}.
The strongest among all E-region instabilities, the Farley-Buneman
instability (FBI), is excited when the relative speed between the average
electron and ion streams exceeds the local ion-acoustic speed. At high
latitudes, this usually occurs when $|\vec{E}_{0}|\gtrsim20$~mV/m.
This and much stronger fields are not
uncommon in the subauroral, auroral, and polar cap areas, especially during
geospace storms and substorms. Driven by magnetospheric activities,
small-scale E-region instabilities heat electrons
\cite<e.g.,>[and references therein]{Foster:Simultaneous00,Milikh:Model03,OppenheimKinetic2013}
and may affect ionospheric conductivities, which in turn
exert large-scale feedback on the magnetosphere
\cite{Merkin:Effect2005GeoRL,DimantCoupling_II_2011JGR,Wiltberger:Effects17}.

At the same time, high-latitude regions are characterized by strong electron
precipitation that gives rise to such spectacular phenomena as Aurora
Borealis. The electron energy distribution in the aurora displays many
different forms that are usually described by qualitative criteria developed
by \citeA{Newell:Diffuse09}. The majority of these
electron energy distributions are classified as diffuse, monoenergetic, or
broadband. In
Section~\ref{Electron Precipitation in Aurora: General Discussion}, we provide
a detailed description of different kinds of aurora.

In this paper, we focus on auroral regions where intense electron
precipitation may overlap with strong driving DC fields. We study the
distribution function modifications caused by precipitating electrons and the
effect of these modifications on the instability development. Using a
physics-based model of electron precipitation from the STET model described below,
we study how this precipitation affects the E-region instability criterion.

We show that precipitating electrons of sufficiently high energies, $E_{0}\gtrsim5$~keV,
can easily penetrate down to the E-region. The plasma turbulence associated
with the E-region instabilities does not have sufficient energy to exert
noticeable feedback on the behavior of precipitating electrons, however this paper shows that the reverse may be quite common. We show this by using the
kinetic linear theory of the FBI and estimating the effect of precipitating
superthermal electrons (SE) on its onset criterion.

The paper is organized as follows. In section~\ref{Background}, we discuss in general
the E-region instabilities (section~\ref{Outline of the Farley-Buneman Instability Onset}),
as well as electron precipitation
and the kinetic tool to treat the latter (section~\ref{Electron Precipitation in Aurora: General Discussion}).
In section~\ref{Results of kinetic simulation}, we present the results of our kinetic simulations using STET.
Section~\ref{Effect of superthermal electrons on the FB instability threshold: kinetic theory} is the central
section of the paper, where we present our theoretical analysis. To obtain a useful insight into the possible effect of
superthermal electrons on the FBI, we start with an oversimplified three-fluid analysis
(section~\ref{Three-fluid linear analysis of the FB instability}). Then, in
section~\ref{Kinetic analysis of the FB instability onset for the general electron distribution function},
we present our principal theoretical analysis based on a hybrid approach
(an approximate kinetic theory for electrons combined with the fluid model of ions). In
section~\ref{Specific calculations for superthermal electrons produced by electron precipitation},
using the results of our kinetic simulations described in section~\ref{Results of kinetic simulation},
we give simple estimates of the modified instability threshold in the presence of superthermal electrons.
In section~\ref{Conclusions}, we give the conclusions and the final discussion. In the appendices, we give some details of analytic
approximations of the collision cross-section and distribution function.

\section{Background\label{Background}}

\subsection{Outline of the FBI
Onset\label{Outline of the Farley-Buneman Instability Onset}}

Here we briefly describe the FBI linear theory relevant to our topic.
All E-region instabilities occur within the lower altitude range of the
ionosphere where electrons are strongly magnetized, $\omega_{ec}\gg\nu_{en}$,
while ions are fully or partially unmagnetized due to their frequent
collisions with neutral atmospheric molecules, $\omega_{ic}\lesssim\nu_{in}$,
where $\omega_{ec}$ and $\omega_{ic}$ are the electron and ion cyclotron
frequencies; $\nu_{en}$ and $\nu_{in}$ are the electron-neutral ($e$-$n$) and
ion-neutral ($i$-$n$) mean collision frequencies, respectively (for
simplicity, we assume only single-species ions). The mean collision
frequencies are altitude-dependent parameters averaged over the entire
particle distributions. To avoid a confusion, we note that in the
kinetic description of electrons (section
\ref{Effect of superthermal electrons on the FB instability threshold: kinetic theory})
we will use the same notation for the velocity-dependent $e$-$n$ collision frequency,
$\nu_{en}(V_{e})$ (the reader should pay attention to the context). At high latitudes, the above conditions
usually hold at E-region altitudes between 90 and 120~km.

All E-region instabilities excite low-frequency plasma-density
compression/\-rarefaction waves. Though such waves are usually acoustic-like,
no long-lived ion acoustic waves can exist in the highly dissipative
E-region ionosphere. Long-lived plasma waves persist there only due to an
external DC electric field, $\vec{E}_{0}\perp\vec{B}$. These plasma waves are
quasineutral, where weak charge separation gives rise to coupled
electrostatic field oscillations. The frequencies of these excited waves,
$\omega$, are typically lower than the characteristic collision frequencies,
$\omega\lesssim\nu_{in}\ll\nu_{en}$, while the typical wavelengths are larger
than the $i$-$n$ collisional mean free path. The corresponding wavevectors,
$\vec{k}$, are largely perpendicular to $\vec{B}$. In the perpendicular to $\vec{B}$ plane, depending on
the driving-field magnitude, $E_{0}$, the wavevectors are mostly directed
within a cone of a few tens of degrees wide around the electron $\vec{E}%
_{0}\times\vec{B}$-drift velocity, $\vec{V}_{0}=\vec{E}_{0}\times\vec{B}%
/B^{2}$, where $B=|\vec{B}|$.

If the particle velocity distributions are close to Maxwellian then, for
sufficiently long-wavelength waves (see below), the
linear stage of the FBI and other E-region
instabilities can be reasonably well described by a closed set of five-moment
fluid-model equations that includes the continuity equation, the momentum, and
the energy balance equations \cite<e.g.,>{Dimant:Ion04}:
\begin{linenomath*}
\begin{subequations}
\label{fluid_model}%
\begin{align}
&  \frac{\partial n_{s}}{\partial t}+\nabla\cdot(n_{s}\vec{V}_{s}%
)=0,\label{continuity_s}\\
&  m_{s}\ \frac{D_{s}\vec{V}_{s}}{Dt}=q_{s}(\vec{E}+\vec{V}_{s}\times\vec
{B})-\frac{\nabla P_{s}}{n}-m_{s}\nu_{sn}\vec{V}_{s},\label{momentum_s}\\
&  n_{s}^{3/2}\ \frac{D_{s}}{Dt}\left(  \frac{T_{s}}{n_{s}^{3/2}}\right)
=\frac{2}{3}\ M_{sn}\nu_{sn}V_{s}^{2}-\delta_{sn}\nu_{s}\left(  T_{s}%
-T_{n}\right)  . \label{energy_s}%
\end{align}
\end{subequations}
\end{linenomath*}
Here the subscript $s=e,i$ characterizes a specific plasma fluid;
$D_{s}/Dt\equiv\partial_{s}/\partial t+\vec{V}_{s}\cdot\nabla$; $\vec{V}_{s}$,
$m_{s}$, $q_{s}$, and $T_{s}$ are the $s$-particle mean flow velocities,
particle masses, charges ($q_{i}=e$, $q_{e}=-e$), and temperatures (in energy
units); $V_{s}=|\vec{V}_{s}|$; $m_{n}$ and $T_{n}$ are the neutral mass and
temperature, respectively; $P_{s}\approx n_{s}T_{s}$ is the $s$-fluid
pressure; $M_{sn}=m_{s}m_{n}/(m_{s}+m_{n})$ is the effective mass of the two
colliding particles ($s$ and $n$); and $\delta_{sn}$ is the average fraction
of energy lost by the particle of the $s$-species during one $s$-$n$ collision;
here $\delta_{in}\simeq1$ and $\delta_{en}\simeq(2$--$4)\times10^{-3}$
\cite{Gurevich:Nonlinear78}. Equation~(\ref{fluid_model}) implies
the frame of reference attached to the mean neutral flow.

In equation~(\ref{fluid_model}), we assume a single ion ($i$) species because
the two major E-region ion components, O$_{2}^{+}$ and NO$_{2}^{+}$, have
fairly close masses ($m_{i}\simeq30~m_{p}$, where $m_{p}$ is the proton mass)
and comparable collision frequencies; the same can be assumed for neutrals
($n$). Note that this simplified set of moment equations misses a few factors,
e.g., ionization-recombination, anisotropic pressure,
viscosity, and heat conductivity, which play little to no role in E-region waves.

For the E-region processes, the two sets of the moment equations for plasma particles
are usually closed through the quasineutrality condition,
$n_{e}\approx n_{i}=n$. This condition presumes sufficiently
long-wavelength waves, compared to the Debye length. This eliminates the
need for Poisson's equation and allows one to unambiguously determine the
turbulent total electric field, $\delta\vec{E}=\vec{E}-\vec{E}_{0}%
=-\nabla\delta\Phi$, where $\delta\Phi$ is the corresponding electrostatic potential.

Wave perturbations of the particle temperatures are crucial for the
aforementioned thermal instabilities, but for the pure FBI the temperatures
$T_{s}$ can be assumed, e.g., constant, $T_{s}\approx T_{s0}$ (the isothermal
regime) or obeying the adiabatic regime, $T_{s}\propto n_{s}^{3/2}$
($P_{s}\propto n_{s}^{5/2}$); the latter is derived by
equation~(\ref{energy_s}) if the right-hand side (RHS) equals zero. Assuming
any of these regimes, the fluid-model description of the pure FBI no longer
requires equation~(\ref{energy_s}).

The electron inertia in the corresponding left-hand side (LHS) of
equation~(\ref{momentum_s}) never plays a role for low-frequency E-region
processes \cite{DimantCoupling_II_2011JGR}. For
inertialess electrons, equation~(\ref{momentum_s}) yields a simple explicit
expression for the electron flow velocity in terms of $\vec{E}$ and $\nabla
P_{e}$:
\begin{linenomath*}
\begin{equation}
\vec{V}_{e}\approx -\ \frac{1}{m_{e}}\left[
\begin{array}
[c]{ccc}%
\nu_{en}/\omega_{ec}^{2} & 1/\omega_{ec} & 0\\
-1/\omega_{ec} & \nu_{en}/\omega_{ec}^{2} & 0\\
0 & 0 & 1/\nu_{en}%
\end{array}
\right]  \times\left[  e\vec{E}+\frac{\nabla P_{e}}{n}\right]  , \label{V_e}%
\end{equation}
\end{linenomath*}
where we used the aforementioned condition of $\omega_{ec}\gg\nu_{en}$. The
3-D vector combination $[\vec{A}]$ on the far right of equation~(\ref{V_e})
implies a right-handed Cartesian coordinate system $[A_{x},A_{y},A_{z}]$ with
the $z$-axis directed along $\vec{B}$. In the preceding $3\times3$ matrix
[$B_{\alpha\beta}$], the two equal diagonal elements, $\nu_{en}/\omega
_{ec}^{2}$, correspond to the electron Pedersen mobility, the remaining
diagonal element, $1/\nu_{en}$, corresponds to the parallel to $\vec{B}$
mobility, whereas the two non-zero non-diagonal elements, $\pm1/\omega_{ec}$,
describe the Hall mobility of the strongly magnetized electrons.
Equation~(\ref{V_e}) applies to both the zero-order background electron flow
velocity, $\vec{V}_{e}$, and wave perturbations, $\delta\vec{V}_{e}$. For
ions, however, the particle inertia in the LHS of equation (\ref{momentum_s})
is crucial for driving the FBI. As a result, the expression for the background
ion flow velocity, $\vec{V}_{i0}$, is analogous to equation~(\ref{V_e}), while
the corresponding wave perturbations, $\delta\vec{V}_{i}$, are described in a
more complex way \cite<see, e.g.,>{DimantCoupling_II_2011JGR}.

In the E-region ionosphere, strongly magnetized electrons move
against the neutral atmosphere with approximately the $\vec{E}_{0}\times
\vec{B}$-drift velocity, $\vec{V}_{e0}\approx\vec{V}_{0}$, while unmagnetized
ions are almost attached to the neutral atmosphere. The background
charged-particle temperatures are usually larger than the neutral temperature,
in part due to the ohmic heating by the driving DC field, $\vec{E}_{0}$. The
background parameters determine the phase velocity of the linearly generated
waves, the linear growth rate, and the threshold-field amplitude for exciting
the FBI, $E_{\mathrm{Thr}}$. The minimum threshold field is usually reached
for sufficiently long-wavelength waves, compared to the ion-neutral ($i$-$n$)
collision mean free path, and for the wavevectors $\vec{k}$ parallel to
$\vec{V}_{0}$. Near the optimum wavevector direction in the perpendicular to
$\vec{B}$ plane, with small but finite $k_{\parallel}\ll k_{\perp}$, the FBI
threshold-field amplitude can be written in the form given, e.g., by
\citeA{Dimant:Model03,Milikh:Model03}:
\begin{linenomath*}
\begin{equation}
E_{\mathrm{Thr}}=(1+\psi)\left(  \frac{1+\kappa_{i}^{2}}{1-\kappa_{i}^{2}%
}\right)  ^{1/2}E_{1}, \label{E_Thr_min}%
\end{equation}
\end{linenomath*}
where $\kappa_{s}=\omega_{sc}/\nu_{s}$ are the magnetization parameters for
the $s$-species and
\begin{linenomath*}
\begin{equation}
\psi=\frac{1}{\kappa_{e}\kappa_{i}}\left(  1+\frac{k_{\parallel}^{2}%
\omega_{ec}^{2}}{k_{\perp}^{2}\nu_{en}^{2}}\right)  . \label{psi_standard}%
\end{equation}
\end{linenomath*}
Equation~(\ref{E_Thr_min}) implies $\kappa_{i}<1$, since above the ion
magnetization boundary, $\kappa_{i}=1$, at high latitudes located around
120~km of altitude, the pure FBI cannot be excited, as stated in
\citeA{Dimant:Ion04}. In the RHS of
equation~(\ref{V_e}), the smallest FBI threshold field, $E_{1}$, corresponding
to altitudes with concurrently small $\psi$ and $\kappa_{i}^{2}$
\cite[Fig.~5]{Dimant:Ion04} is given by
\begin{linenomath*}
\begin{equation}
E_{1}=C_{s}B=20\left(  \frac{T_{e}+T_{i}}{600~\mathrm{K}}\right)
^{1/2}\left(  \frac{B}{5\times10^{4}\mathrm{nT}}\right)  \mathrm{mV}%
/\mathrm{m},%
 \label{E_1}
\end{equation}
\end{linenomath*}
where $C_{s}=[(T_{e}+T_{i})/m_{i}]^{1/2}$ is the isothermal ion-acoustic
speed. For the adiabatic regime of instability generation, one must replace
$T_{s}$ with $(5/3)T_{s}$.

Equations~(\ref{fluid_model})--(\ref{E_1}) hold for the particle velocity
distributions that are reasonably close to Maxwellian. However, during strong
diffuse or discrete aurora, the electron distribution function changes
dramatically: in addition to the nearly Maxwellian cold-temperature thermal
bulk with energies well below 0.1~eV, a significant superthermal tail develops
within the eV-to-tens of keV energy range, as we discuss in more detail in the
following sections. With significant non-Maxwellian additions to the velocity
distribution, approximate fluid-model equations (\ref{fluid_model})-(\ref{V_e})
lose their validity, as we demonstrate below in
section~\ref{Effect of superthermal electrons on the FB instability threshold: kinetic theory}.
Processes with non-Maxwellian velocity distributions require the kinetic description.

For the kinetic treatment, of significant importance is the fact that after
$e$-$n$ collisions most electrons change their momentum at a much
higher rate than they lose their energy. This means that $e$-$n$ collisions
effectively scatter electrons by large angles in the velocity space
with only small relative changes in their kinetic energies. As a result, the
electron distribution function remains almost isotropic, $f_{e}(\vec
{V})\approx F_{0}(V)$ (here $V\equiv|\vec{V}_{e}|$), but $F_{0}(V)$ may deviate significantly from the
Maxwellian velocity distribution. Due to this effective isotropization, when calculating integral scalar quantities like the
local electron density or pressure, instead of the general 3-D velocity
integration one can use a much simpler 1-D speed integration, $\int%
(\cdots)f_{0}(\vec{V})d^{3}V\approx4\pi\int_{0}^{\infty}(\cdots)F_{0}%
(V)V^{2}dV$.

In particular, the total electron density, pressure, and temperature, used in
fluid-model equation~(\ref{fluid_model}), become:
\begin{linenomath*}
\begin{subequations}
\label{n,P,T_tot}%
\begin{align}
n_{e}  &  \approx4\pi\int_{0}^{\infty}F_{0}(V)V^{2}dV,\label{n_tot}\\
P_{e}  &  =n_{e}T_{\mathrm{tot}}\approx\frac{4\pi m_{e}}{3}\int_{0}^{\infty
}F_{0}(V)V^{4}dV,\label{Pe}\\
T_{\mathrm{tot}}  &  \approx\frac{4\pi m_{e}}{3n_{0}}\int_{0}^{\infty}%
F_{0}(V)V^{4}dV=\frac{m_{e}\int_{0}^{\infty}F_{0}(V)V^{4}dV}{3\int_{0}%
^{\infty}F_{0}(V)V^{2}dV}. \label{T_tot}%
\end{align}
\end{subequations}
\end{linenomath*}
It is important that the dominant omnidirectional part of the electron
distribution function, $F_{0}(V)$, includes both the thermal bulk and
superthermal tail. As we demonstrate below, the relative addition of
superthermal particles to the total electron density, $n_{e}$, is usually
small and can be neglected, $n_{e}\approx n_{0}$, while the total
temperature, $T_{\mathrm{tot}}$, due to the additional multiplier $V^{2}$ in
the integrand of (\ref{T_tot}), can exceed the electron bulk temperature,
$T_{e0}$, dramatically. Note that in the high-latitude nighttime E-region
ionosphere a significant fraction of the thermal bulk plasma may originate from the
electron precipitation followed by ionizing collisions of the precipitated
energetic electrons. However, this happens only after multiple collisions
causing the electrons to have already cooled down and
become the effectively `maxwellized' distribution within the cold-temperature
thermal bulk. This cold plasma is redistributed by drifts between different locations and can
survive without the local precipitation source for a sufficiently long time.

A naive viewpoint suggests using the modified temperature given by
equation~(\ref{T_tot}) to determine the modified FBI threshold by merely
replacing in equation~(\ref{E_1}) the regular electron temperature $T_{0}$
with $T_{\mathrm{tot}}$. However, the analysis of
section~\ref{Effect of superthermal electrons on the FB instability threshold: kinetic theory} below
shows that the velocity dependence of the $e$-$n$ collision frequency of
electrons makes this approach inaccurate.

\subsection{Electron Precipitation and STET code\label{Electron Precipitation in Aurora: General Discussion}}

The electron energy distribution in the aurora displays many different forms
that are usually described by qualitative criteria developed by
\citeA{Newell:Diffuse09,Newell:Substorm10}. The
majority of these electron distributions are classified as diffuse,
monoenergetic, or broadband. Details of these characterization criteria and
the origin of different class of electron precipitation phenomena were
discussed by \citeA{McIntosh:Maps14} and will not
be repeated here.

The diffuse aurora is primarily caused by wave-particle interactions of high
energy electrons, $\mathcal{E}\equiv m_{e}V^{2}/2>1$ keV, within the plasma sheet
\cite{Thorne:Scattering10}. Note that since in this paper we treat only electrons kinetically, we will drop
subscripts $e$ from any kinetic characteristics (like $V$).

The non-steady state SuperThermal Electron Transport (STET) code,
used in this project, was initially developed by
\citeA{Khazanov:Non-steady-state93} for SE transport in
the plasmasphere. Later, this code was further generalized for MI coupling
studies in the region of diffuse and monoenergetic auroras by
\citeA{Khazanov:Magnetosphere14,Khazanov:Role16,Khazanov:Ionosphere16,%
Khazanov:Major17,Khazanov:Magnetosphere21} and validated experimentally
by \citeA{Samara:First17} in their case study of a
pulsating auroral event imaged optically at high time resolution. The results
of our simulation were also successfully compared to FAST
\cite{Khazanov:Role16,Khazanov:Ionosphere16} and
DMSP \cite{Khazanov:Electron21} observations.

There are different settings available for the application of the STET model
for studying the diffuse aurora. The first setting involves imposing a
spectrum of primary precipitating electrons with energies above 500--600 eV at
an altitude of 800 km and keeping the spectra unchanged
\cite{Khazanov:Ionosphere16}. This setting implicitly
assumes no MI coupling processes for the energy range of the imposed
precipitation and implies the usage of experimental energy fluxes as the
boundary conditions. \citeA{Khazanov:Ionosphere16} introduced a modification of this boundary
condition setting to account for the role of multiple reflections
(backscatters) of degraded primary electrons traveling between two
magnetically conjugate hemispheres. The latest setting is adapted in the
studies that are presented below.

The STET setting in the region of monoenergetic aurora is similar, but assumes
the existence of the electrostatic acceleration region located at altitudes of
$(1.5$--$2)R_{E}$ \cite{Marklund:Altitude11}, where $R_{E}$ is
the Earth's radius. Specifically, STET code was set up to run from the
northern or southern hemispheres from 90 km to $2R_{E}$, where $R_{E}$ is the
Earth's radius, with the multiple reflection (backscatter) of the electrons
whose electrostatic energies are smaller than the potential drop of the
acceleration region.

In the analysis presented below, in order to describe the primary
magnetosphere-driven electron precipitation in the region of diffuse aurora,
we used only the Maxwellian EDF input in the energy range of 600~eV to 30~keV,
\begin{linenomath*}
\begin{equation}
\Phi(\mathcal{E})=C\mathcal{E}e^{-\mathcal{E}/\mathcal{E}_{0}},
\label{Maxwellian}%
\end{equation}
\end{linenomath*}
where $\Phi
=2\mathcal{E}f_{e}(\vec{V},s,t)
/m^{2}$ is the 
SE flux \cite{Khazanov:Kinetic11}, $\mathcal{E}_{0}$ is the characteristic energy of plasmasheet electrons,
and $C$ is the normalization constant for the selection of the integrated
energy flux driven by magnetospheric processes. In the region of monoenergetic
aurora, we selected the Gaussian distribution, as in \citeA{Banks:New74},
\begin{linenomath*}
\begin{equation}
\Phi(\mathcal{E})=A\exp[-(\mathcal{E}-\mathcal{E}_{0})^{2}%
/(2\sigma^{2})],\qquad\sigma=0.1\mathcal{E}_{0}. \label{Phi_Gaussian}%
\end{equation}
\end{linenomath*}
Here $\mathcal{E}_{0}$ is also the characteristic energy of monoenergetic
accelerated electrons and $A$ is the normalization constant for the selection of
the integrated energy flux as defined above, but only for the electrostatic
acceleration region.

In accord with equation (\ref{n,P,T_tot}), the density and pressure of the
superthermal electron population, were
found using
\begin{linenomath*}
\begin{subequations}
\label{nP_SE}%
\begin{align}
n_{\mathrm{SE}}  &  =4\pi\int_{\mathcal{E}_{\min}}^{\mathcal{E}_{\max}}%
\frac{\Phi_{0}}{V(\mathcal{E})}\ d\mathcal{E},\label{n_SE}\\
P_{\mathrm{SE}}  &  =\frac{4\pi}{3}\int_{\mathcal{E}_{\min}}^{\mathcal{E}%
_{\max}}\frac{\mathcal{E}\Phi_{0}}{V(\mathcal{E})}\ d\mathcal{E}, \label{P_SE}%
\end{align}
\end{subequations}
\end{linenomath*}
where $V(\mathcal{E})=(2\mathcal{E}/m_{e})^{1/2}$.
The minimum and maximum superthermal energies used in our simulations were
$\mathcal{E}_{\min}=1$~eV and $\mathcal{E}_{\max}=30$~keV, respectively.

These values of $n_{\mathrm{SE}}$ and $P_{\mathrm{SE}}$ are calculated below
for different types of precipitated electron spectra of
equations~(\ref{Maxwellian}) and (\ref{Phi_Gaussian}), modeling the diffuse
and monoenergetic auroras, respectively. We used the following inputs into the
STET model. The neutral thermospheric densities and temperatures were given by
MSIS-90 \cite{Hedin:Extension91}. The electron profile in the
ionosphere was calculated based on the IRI model
\cite{Bilitza:International17} and extended into the
magnetosphere under the assumption that the electron thermal density
distribution in the magnetosphere is proportional to the geomagnetic field as
$n_{e}\propto B^{1/2}$. Cross-sections for elastic collisions, state-specific
excitation, and ionization were taken from \citeA{Solomon:Auroral88}.

\section{Results of Kinetic Simulation\label{Results of kinetic simulation}}

Before presenting the simulation results, we notice the following. In our
simulations, we neglect any collisions between the superthermal particles
themselves, compared to their collisions with the thermal bulk particles.
Furthermore, in the lower ionosphere the $e$-$n$ collisions vastly dominate
over Coulomb collisions between the charged particles. With neglect of
electron-electron collisions, the corresponding kinetic equation becomes linear with
respect to the superthermal particle flux $\Phi_0$. As a result, given the
energy distribution of the precipitated electrons, any superthermal
particle-energy-integrated characteristics, such as the density and pressure,
will be proportional to the mean energy flux, $\Phi_{\mathcal{E}}$. The latter
is defined as the total SE energy (in ergs) per unit square (in cm$^{2}$) per
unit time (in s) at a given altitude. We will represent the SE density and
pressure defined by equation (\ref{nP_SE}) as
\begin{linenomath*}
\begin{equation}
n_{\mathrm{SE}}=A_{n}\left(  \frac{\Phi_{\mathcal{E}}}{10~\mathrm{erg}%
~\mathrm{cm}^{-2}~\mathrm{s}^{-1}}\right)  ,\qquad P_{\mathrm{SE}}%
=A_{P}\left(  \frac{\Phi_{\mathcal{E}}}{10~\mathrm{erg}~\mathrm{cm}%
^{-2}~\mathrm{s}^{-1}}\right)  .\label{n_SE,P_SE}%
\end{equation}
\end{linenomath*}
The characteristic SE density and pressure values, $A_{n,P}$, corresponding to
the moderately high energy-flux value of $10~\mathrm{erg}~\mathrm{cm}%
^{-2}~\mathrm{s}^{-1}$, depend on the SE velocity distribution and the
ionosphere-thermosphere parameters at given ionospheric altitudes. For our
simulations, we picked three E-region altitudes: 100, 110, and 120~km, that
best characterize the typical altitude range for the FBI generation.

We performed specific kinetic simulations for the following conditions.
Bearing in mind both the discrete and diffuse aurorae, we modeled the
Maxwellian EDF given by equation (\ref{Maxwellian}) and the Gaussian flux
given by equation~(\ref{Phi_Gaussian}). To characterize various precipitation
conditions, we have chosen different values of $\mathcal{E}_{0}$ for each EDF
(see below). We believe that these values are most characteristic for each
kind of storm-time EDF.

To compare the SE contributions to the total electron density and pressure, we
have chosen the following values of the nighttime cold thermal background from
the IRI model, as shown in Table~\ref{Table:IRI}.
\begin{linenomath*}
\begin{table}
\caption{Nighttime cold thermal background from the IRI model}
\centering
\begin{tabular}
[c]{|l|l|l|}\hline
Altitude, km & Background density, cm$^{-3}$ & Background pressure,
eV~cm$^{-3}$\\\hline
100 & $1.67\times10^{3}$ & 53.1\\\hline
110 & $2.59\times10^{3}$ & 82.4\\\hline
120 & $1.08\times10^{3}$ & 34.2\\\hline
\end{tabular}
\label{Table:IRI}
\end{table}
\end{linenomath*}
The background electron temperature for all three altitudes was taken equal,
$T_{e0}=370$~K (corresponding to 0.0318~eV).

\begin{figure}
\noindent\includegraphics[width=\textwidth]{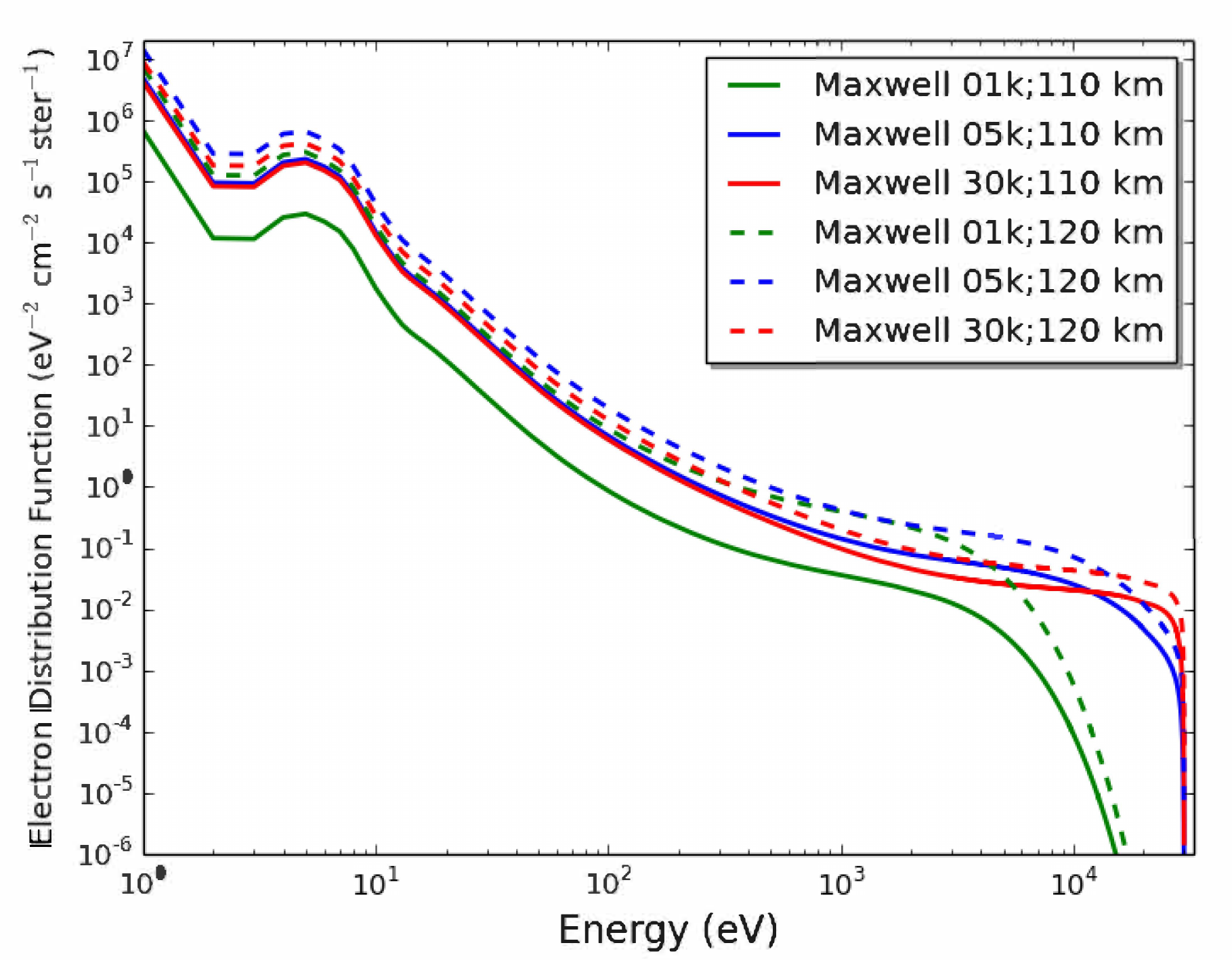}
\caption{Superthermal electron energy distribution function: some results of kinetic (STET) simulations (Maxwell input); the parameters are shown in the figure.}
\label{Fig:STET}
\end{figure}
Figure~\ref{Fig:STET} shows examples of the corresponding SE velocity distributions. Tables~\ref{Table:Maxwellian_1keV}-\ref{Table:Maxwellian_30keV} show some simulation results for the Maxwell-input EDF. Tables~\ref{Table:Gaussian_1keV}-\ref{Table:Gaussian_30keV} show some simulation results for the Gaussian EDF.

\begin{linenomath*}
\begin{table}
\renewcommand\thetable{1a}
\caption{Maxwellian EDF, $\mathcal{E}_{0}=1$~keV}
\centering
\begin{tabular}
[c]{|l|l|l|}\hline
Altitude, km & $A_{n}$, cm$^{-3}$ & $A_{P}$, eV~cm$^{-3}$\\\hline
100 & 2.06 & 18.6\\\hline
110 & 21 & 180\\\hline
120 & 82 & 635\\\hline
\end{tabular}
\label{Table:Maxwellian_1keV}
\end{table}
\end{linenomath*}
\begin{linenomath*}
\begin{table}
\renewcommand\thetable{1b}
\caption{Maxwellian EDF, $\mathcal{E}_{0}=10$~keV}
\centering
\begin{tabular}
[c]{|l|l|l|}\hline
Altitude, km & $A_{n}$, cm$^{-3}$ & $A_{P}$, eV~cm$^{-3}$\\\hline
100 & 10 & 604\\\hline
110 & 23 & 1441\\\hline
120 & 37 & 2195\\\hline
\end{tabular}
\label{Table:Maxwellian_10keV}
\end{table}
\end{linenomath*}
\begin{linenomath*}
\begin{table}
\renewcommand\thetable{1c}
\caption{Maxwellian EDF, $\mathcal{E}_{0}=30$~keV}
\centering
\begin{tabular}
[c]{|l|l|l|}\hline
Altitude, km & $A_{n}$, cm$^{-3}$ & $A_{P}$, eV~cm$^{-3}$\\\hline
100 & 13.29 & 1001\\\hline
110 & 28.2 & 2232\\\hline
120 & 41.1 & 3255\\\hline
\end{tabular}
\label{Table:Maxwellian_30keV}
\end{table}
\end{linenomath*}

\begin{linenomath*}
\begin{table}
\renewcommand\thetable{2a}
\caption{Gaussian EDF, $\mathcal{E}_{0}=1$~keV}
\centering
\begin{tabular}
[c]{|l|l|l|}\hline
Altitude, km & $A_{n}$, cm$^{-3}$ & $A_{P}$, eV~cm$^{-3}$\\\hline
100 & 0.07 & $4.81\times10^{-2}$\\\hline
110 & 4.45 & 3.16\\\hline
120 & 55.0 & 41.6\\\hline
\end{tabular}
\label{Table:Gaussian_1keV}
\end{table}
\end{linenomath*}
\begin{linenomath*}
\begin{table}
\renewcommand\thetable{2b}
\caption{Gaussian EDF, $\mathcal{E}_{0}=10$~keV}
\centering
\begin{tabular}
[c]{|l|l|l|}\hline
Altitude, km & $A_{n}$, cm$^{-3}$ & $A_{P}$, eV~cm$^{-3}$\\\hline
100 & 6.72 & 158.8\\\hline
110 & 18.56 & 586.4\\\hline
120 & 28.41 & 1047\\\hline
\end{tabular}
\label{Table:Gaussian_10keV}
\end{table}
\end{linenomath*}
\begin{linenomath*}
\begin{table}
\renewcommand\thetable{2c}
\caption{Gaussian EDF, $\mathcal{E}_{0}=30$~keV}
\centering
\begin{tabular}
[c]{|l|l|l|}\hline
Altitude, km & $A_{n}$, cm$^{-3}$ & $A_{P}$, eV~cm$^{-3}$\\\hline
100 & 6.94 & 1066\\\hline
110 & 12.37 & 2099\\\hline
120 & 15.98 & 2823\\\hline
\end{tabular}
\label{Table:Gaussian_30keV}
\end{table}
\end{linenomath*}

Using the table values along with equation (\ref{n_SE,P_SE}) and comparing the
simulation results with the typical background parameters, we see that the
contribution of the SE tail, $n_{\mathrm{SE}}$, to the total electron density,
$n_{\mathrm{tot}}=n_{e0}+n_{\mathrm{SE}}$, is usually small compared to the
background density, $n_{e0}$. For reasonable values of the mean energy flux,
$\Phi_{\mathcal{E}}\lesssim10$ erg~cm$^{-2}$ s$^{-1}$, even for the large
values of $\mathcal{E}_{0}$, the SE contribution to $n_{\mathrm{tot}}$ can be neglected.

An entirely different situation, however, takes place for the total pressure,
$P_{\mathrm{tot}}$, and hence for the total electron temperature,
$P_{\mathrm{tot}}=P_{e0}+P_{\mathrm{SE}}$. Only for the Maxwellian EDF with
the smallest calculated SE energy $\mathcal{E}_{0}=1$~keV the SE contributions
is comparable to the background values of the electron pressure and
temperature; for all other values of $\mathcal{E}_{0}$, $\Phi_{\mathcal{E}%
}\gtrsim10$ erg cm$^{-2}$ s$^{-1}$, and both kinds of the EDF the SE
contributions are much larger than the corresponding background values.

\section{Effect of Superthermal Electrons on the FBI
Threshold: analytical treatment\label{Effect of superthermal electrons on the FB instability threshold: kinetic theory}}

As we have already mentioned, the naive calculation of the modified
instability threshold based on replacing the undisturbed electron cold-plasma
temperature $T_{e}$ with the modified temperature $T_{\mathrm{tot}}$ defined
by equation~(\ref{n,P,T_tot}) turns out to be incorrect. This will become
clear after we implement in section~\ref{Three-fluid linear analysis of the FB instability}
a tentative three-fluid approach based on two distinct Maxwellian velocity distributions
of electrons (the cold thermal bulk and the energetic tail of precipitated electrons) with two
different $e$-$n$ collision frequencies.

At the E-region altitudes, however, the energy distribution of precipitating
electrons deviates significantly from a Maxwellian distribution. Also, the
$e$-$n$ collision frequency depends smoothly on the electron energy and hence
cannot be reduced to only two constant values. All this requires the proper
description of the electron behavior to be kinetic. At the same time, for
sufficiently long-wavelength and low-frequency waves, as specified below by
equation~(\ref{k_conditions}), the ion behavior can be successfully described
by the much simpler fluid model.

For the kinetic treatment of electrons, we will mostly follow the approximate
kinetic approach developed in \citeA{Dimant:Kinetic95a}. This approach is based
on the assumption that $e$-$n$ collisions lead to much faster angular scatter
of electrons in the velocity space than to losses of their energies, as we
already mentioned in section~\ref{Outline of the Farley-Buneman Instability Onset}. For superthermal
electrons with energies $\mathcal{E}\gtrsim1~$keV, the two rates are comparable, but the
approximate approach of \citeA{Dimant:Kinetic95a} is still useful and will
lead to reasonably accurate analytic results.

\subsection{Three-Fluid Linear Analysis of the
FBI\label{Three-fluid linear analysis of the FB instability}}

In this section, we outline the simplified 3-fluid model approach by assuming
two electron fluids and a single ion one. This oversimplified approach does
not rival the rigorous kinetic approach implemented in the following
section~\ref{Kinetic analysis of the FB instability onset for the general electron distribution function},
but it will provide useful insight into the effect of electron precipitation
on the FBI threshold conditions and will help identify the key factors.

The closed set of fluid-model equations~(\ref{fluid_model}) is only valid for
the particle velocity distributions reasonably close to Maxwellian, so that in
this tentative approach we will use the model of two Maxwellian populations of
electrons, $f_{e}=f_{\mathrm{TB}}+f_{\mathrm{SE}}$, where TB stands for the
thermal bulk and SE stands for the superthermal electron tail. Each Maxwellian
population has its own density and temperature: $f_{\mathrm{TB}}%
=n_{\mathrm{TB}}[m_{e}/(2\pi T_{\mathrm{TB}})]^{3/2}\exp(-\mathcal{E}%
/T_{\mathrm{TB}})$, and $f_{\mathrm{SE}}=n_{\mathrm{SE}}[m_{e}/(2\pi
T_{\mathrm{SE}})]^{3/2}\exp(-\mathcal{E}/T_{\mathrm{SE}})$. Under the actual
conditions of electron precipitation, the conditions $T_{\mathrm{SE}}\gg
T_{\mathrm{TB}}$, $n_{\mathrm{SE}}$, and $n_{\mathrm{TB}}$ usually hold,
although for this specific treatment these conditions are of no importance and
will not be imposed.

Adding to the two electron fluids an ion fluid and assuming for simplicity the isothermal regime of the
pure FBI with constant $T_{\mathrm{TB}}$, $T_{\mathrm{SE}}$, and $T_{i}$ but variable
densities, $n_{\mathrm{SE}}$, $n_{\mathrm{TB}}\approx n_{i}$, we will need only two first
fluid-model equations~(\ref{continuity_s}) and (\ref{momentum_s}),
\begin{linenomath*}
\begin{subequations}
\label{moments_two}%
\begin{align}
\frac{\partial n_{j}}{\partial t}+\nabla\cdot(n_{j}\vec{V}_{j})  &
=0,\label{my_fluid_equations_cont}\\
m_{j}\left(  \frac{\partial}{\partial t}+\vec{V}_{j}\cdot\nabla\right)
\vec{V}_{j}  &  =q_{j}(\vec{E}+\vec{V}_{j}\times\vec{B})-T_{j}\ \frac{\nabla
n_{j}}{n_{j}}-m_{j}\nu_{j}\vec{V}_{j}, \label{my_fluid_equations_mom}%
\end{align}
\end{subequations}
\end{linenomath*}
where the subscript $j$ denotes either each of the two electron species,
$j=\mathrm{TB}$ and $j=\mathrm{SE}$, or the single ion species, $j=i$. The
kinetic $e$-$n$ collision frequency $\nu_{e}$ depends strongly on the
individual electron velocity. To mimic this in our oversimplified three-fluid
model, we will assign for each electron population its own constant value of
$\nu_{e}$: \ the mean thermal bulk value, $\nu_{\mathrm{TB}}$, and the mean
superthermal value, $\nu_{\mathrm{SE}}$. These values of $\nu_{e}$ may be
vastly different.

Bearing in mind the long-wavelength and low-frequency E-region processes, we will
close all three sets of fluid equations by the quasineutrality condition,
$n_{i}=n_{\mathrm{TB}}+n_{\mathrm{SE}}$, for both the undisturbed plasma
background, $n_{i}^{(0)}=n_{\mathrm{TB}}^{(0)}+n_{\mathrm{SE}}^{(0)}$, and
linear wave perturbations, $\delta n_{i}=\delta n_{\mathrm{TB}}+\delta
n_{\mathrm{SE}}$. In what follows, we will mostly operate with the relative
fractions of each background electron population, $\rho_{\mathrm{TB}%
}=n_{\mathrm{TB}}^{(0)}/n_{i}^{(0)}$ and $\rho_{\mathrm{SE}}=n_{\mathrm{SE}%
}^{(0)}/n_{i}^{(0)}$, so that $\rho_{\mathrm{TB}}+\rho_{\mathrm{SE}}=1$.

For the undisturbed background flows, after setting $\partial/\partial t\rightarrow0$,
$\nabla\rightarrow0$, equation~(\ref{momentum_s}) yields%
\begin{linenomath*}
\begin{equation}
\vec{V}_{j0}=\left.  \left(  \frac{q_{j}\vec{E}_{0}}{m_{j}\nu_{j}}+\kappa
_{j}^{2}\vec{V}_{0}\right)  \right/  \left(  1+\kappa_{j}^{2}\right)  ,
\label{V_s0}%
\end{equation}
\end{linenomath*}
where the $\vec{E}_{0}\times\vec{B}$-drift velocity $\vec{V}_{0}$ and
magnetization parameters $\kappa_{j}=\omega_{cj}/\nu_{j}$ were defined in
section~\ref{Outline of the Farley-Buneman Instability Onset}; $q_{i}=e$,
$q_{\mathrm{TB}}=q_{\mathrm{SE}}=-e$; and $m_{\mathrm{TB}}=m_{\mathrm{SE}}=m_{e}$.
In spite of the common value of the electron gyrofrequency $\omega_{ce}$, each
Maxwellian electron population, TB and SE, has different magnetization
parameters $\kappa_{j}$, $\kappa_{\mathrm{TB}}=\omega_{ce}/\nu_{\mathrm{TB}%
}\neq\kappa_{\mathrm{SE}}=\omega_{ce}/\nu_{\mathrm{SE}}$. In what follows, we
will assume both electron fluids to be strongly magnetized, $\kappa
_{\mathrm{TB}},\kappa_{\mathrm{SE}}\gg1$, so that the average background flow
of all electrons is close to the $\vec{E}_{0}\times\vec{B}$-drift velocity,
$\vec{V}_{\mathrm{TB}0}\approx\vec{V}_{\mathrm{SE}0}\approx\vec{V}_{0}$. For
the following analysis, it is also convenient to introduce the velocity
difference between the background electron and ion flows:
\begin{linenomath*}
\begin{equation}
\vec{U}\approx\vec{V}_{0}-\vec{V}_{i0}=\left.  \left(  \vec{V}_{0}-\frac
{q_{i}\vec{E}_{0}}{m_{i}\nu_{i}}\right)  \right/  \left(  1+\kappa_{i}%
^{2}\right)  . \label{UU}%
\end{equation}
\end{linenomath*}

In all low-frequency E-region processes, electric fields are electrostatic,
$\vec{E}=-\nabla\Phi$, while the magnetic field $\vec{B}$ remains essentially
constant. For linear wave perturbations of all space/time-varying quantities,
we will set the standard harmonic-wave ansatz: $\delta A\propto\exp[i(\vec
{k}\cdot\vec{r}-\omega t)]$ with real $\vec{k}$, but complex frequency,
$\omega=\omega_{r}+i\gamma$.

Introducing dimensionless variables for each species $j$:
\begin{linenomath*}
\begin{equation}
\eta_{j}\equiv\frac{\delta n_{j}}{n_{j0}},\qquad\phi\equiv\frac{e\delta\Phi
}{T_{i0}}, \label{Introducing}%
\end{equation}
\end{linenomath*}
we obtain from continuity equation~(\ref{continuity_s}) a simple relation:
\begin{linenomath*}
\begin{equation}
\eta_{j}=\frac{\vec{k}\cdot\delta\vec{V}_{j}}{\Omega_{j}},%
\label{eta_s}
\end{equation}
\end{linenomath*}
where $\Omega_{j}\equiv\omega-\vec{k}\cdot\vec{V}_{j0}$ is the
Doppler-shifted wave frequency in the frame of
reference of the $j$-species mean flow. The $j$-fluid velocity perturbation
$\delta\vec{V}_{j}$ should be found from momentum-balance
equation~(\ref{momentum_s}). In the dimensionless variables, all $\delta
\vec{V}_{j}$ become proportional to the normalized linear combinations of the
linearized wave electric field with the particle pressure perturbations,
$\alpha_{j}\phi+\eta_{j}$, where
\begin{linenomath*}
\begin{equation}
\alpha_{i}=1,\qquad\alpha_{\mathrm{TB}}=-\ \frac{T_{i0}}{T_{\mathrm{TB}}%
},\qquad\alpha_{\mathrm{SE}}=-\ \frac{T_{i0}}{T_{\mathrm{SE}}}.
\label{alpha_j}%
\end{equation}
\end{linenomath*}
Then equation~(\ref{eta_s}) yields $\eta_{j}$ in terms of $\phi$,
\begin{linenomath*}
\begin{equation}
\eta_{j}=\frac{\alpha_{j}A_{j}}{1-A_{j}}\ \phi,\qquad A_{j}\equiv\frac{\vec
{k}\cdot\delta\vec{V}_{j}}{\left(  \alpha_{j}\phi+\eta_{j}\right)  \Omega_{j}%
}, \label{1}%
\end{equation}
\end{linenomath*}
via still undetermined coefficients $A_{j}$. In the direction parallel to
$\vec{B}$, equation~(\ref{momentum_s}) yields
\begin{linenomath*}
\begin{equation}
\delta\vec{V}_{j\parallel}=-i\ \frac{\vec{k}_{\parallel}V_{Tj}^{2}\left(
\alpha_{j}\phi+\eta_{j}\right)  }{\nu_{j}\left(  1-i\Omega_{j}/\nu_{j}\right)
}, \label{V_s||}%
\end{equation}
\end{linenomath*}
while in the perpendicular to $\vec{B}$ component of $\delta\vec{V}_{j}$, we
have a more complicated relation,
\begin{linenomath*}
\begin{equation}
\delta\vec{V}_{j\perp}=-i\ \frac{V_{Tj}^{2}}{\nu_{j}}\ \frac{\left(
1-i\Omega_{j}/\nu_{j}\right)  \vec{k}_{\perp}+\kappa_{j}(\vec{k}_{\perp}%
\times\hat{b})}{\left(  1-i\Omega_{j}/\nu_{j}\right)  ^{2}+\kappa_{j}^{2}%
}\left(  \alpha_{j}\phi+\eta_{j}\right)  , \label{V_sperp}%
\end{equation}
\end{linenomath*}
where $\vec{k}_{\parallel}$ and $\vec{k}_{\perp}$ are the wavevector
components in the parallel and perpendicular to $\vec{B}$ directions,
respectively. For the coefficients $A_{j}$, these equations yield
\begin{linenomath*}
\begin{equation}
A_{j}=-i\ \frac{V_{Tj}^{2}}{\nu_{j}\Omega_{j}}\left[  \frac{\left(
1-i\Omega_{j}/\nu_{j}\right)  k_{\perp}^{2}}{\left(  1-i\Omega_{j}/\nu
_{j}\right)  ^{2}+\kappa_{j}^{2}}+\frac{k_{\parallel}^{2}}{1-i\Omega_{j}%
/\nu_{j}}\right]  . \label{A_s_opiat}%
\end{equation}
\end{linenomath*}

Using the quasineutrality condition for the wave perturbations, $\eta_{i}%
=\rho_{\mathrm{TB}}\eta_{\mathrm{TB}}+\rho_{\mathrm{SE}}\eta_{\mathrm{SE}}$,
we obtain from equations~(\ref{1}) and (\ref{A_s_opiat}) the three-fluid FBI
dispersion relation:
\begin{linenomath*}
\begin{equation}
D(\omega,\vec{k})\equiv1+\frac{1-A_{i}}{A_{i}}\left(  \frac{\rho_{\mathrm{TB}%
}|\alpha_{\mathrm{TB}}|A_{\mathrm{TB}}}{1-A_{\mathrm{TB}}}+\frac
{\rho_{\mathrm{SE}}|\alpha_{\mathrm{SE}}|A_{\mathrm{SE}}}{1-A_{\mathrm{SE}}%
}\right)  =0, \label{disper_gen}%
\end{equation}
\end{linenomath*}
where we have used the fact that both $\alpha_{\mathrm{TB}}$ and
$\alpha_{\mathrm{SE}}$ are negative, as seen from equation~(\ref{alpha_j}).

General three-fluid FBI dispersion relation, equation~(\ref{disper_gen}), does not have a
simple general solution. Fortunately, it can be simplified by taking into
account the fact that fluid-model equation~(\ref{fluid_model}) is valid only
in the long-wavelength limit in which all wave vectors are much larger than
the corresponding ion collisional mean free paths, while the wave frequencies
are small compared to the ion-neutral collision frequencies, $\left\vert
\omega\right\vert ,\ kV_{i0},kV_{Ti}\ll\nu_{i}$. Otherwise, ion Landau damping
becomes crucial, requiring the kinetic treatment of ions. It is also important
that the minimum values of the FBI threshold field are always reached in the
same long-wavelength limit, where we automatically obtain $\left\vert
A_{j}\right\vert \ll1$. Assuming also $\left\vert \Omega_{\mathrm{TB}}%
/\nu_{\mathrm{TB}}\right\vert $,$\left\vert \Omega_{\mathrm{SE}}%
/\nu_{\mathrm{SE}}\right\vert \ll\left\vert \Omega_{i}/\nu_{i}\right\vert $,
$k_{\parallel}^{2}\ll k_{\perp}^{2}$, and $\kappa_{i}\lesssim1$, we obtain for
all $A_{j}$ simpler expressions:
\begin{linenomath*}
\begin{subequations}
\label{A_e,A_s}%
\begin{align}
A_{\mathrm{TB}}  &  \approx-i\ \frac{k_{\perp}^{2}V_{T\mathrm{TB}}%
^{2}(1+\kappa_{\mathrm{TB}}^{2}k_{\parallel}^{2}/k_{\perp}^{2})}%
{\nu_{\mathrm{TB}}\Omega_{\mathrm{TB}}\kappa_{\mathrm{TB}}^{2}(1-i\Omega
_{\mathrm{TB}}/\nu_{\mathrm{TB}})},\label{A_e}\\
A_{\mathrm{SE}}  &  \approx-i\ \frac{k_{\perp}^{2}V_{T\mathrm{SE}}%
^{2}(1+\kappa_{\mathrm{SE}}^{2}k_{\parallel}^{2}/k_{\perp}^{2})}%
{\nu_{\mathrm{SE}}\Omega_{\mathrm{SE}}\kappa_{\mathrm{SE}}^{2}(1-i\Omega
_{\mathrm{SE}}/\nu_{\mathrm{SE}})},\label{A_s}\\
A_{i}  &  \approx-i\ \frac{k_{\perp}^{2}V_{Ti}^{2}(1-i\Omega_{i}/\nu_{i})}%
{\nu_{i}\Omega_{i}[(1-i\Omega_{i}/\nu_{i})^{2}+\kappa_{i}^{2}]}. \label{A_i}%
\end{align}
\end{subequations}
\end{linenomath*}
Then, to the first-order accuracy with respect to small $\left\vert
A_{j}\right\vert $, general three-fluid FBI dispersion equation~(\ref{disper_gen})
reduces to
\begin{linenomath*}
\begin{align}
D(\omega,\vec{k})  &  \approx1+\frac{\rho_{\mathrm{TB}}|\alpha_{\mathrm{TB}%
}|A_{\mathrm{TB}}}{A_{i}}\left(  1+A_{\mathrm{TB}}-A_{i}\right) \nonumber\\
&  +\frac{\rho_{\mathrm{SE}}|\alpha_{\mathrm{SE}}|A_{\mathrm{SE}}}{A_{i}%
}\left(  1+A_{\mathrm{SE}}-A_{i}\right)  =0. \label{D_reduced}%
\end{align}
\end{linenomath*}
This reduced dispersion relation has certain advantages over general
equation~(\ref{disper_gen}). First, in the assumed long-wavelength limit,
$|\operatorname{Im}D(\omega,\vec{k})|$ turns out to be automatically small
compared to $|\operatorname{Re}D(\omega,\vec{k})|$, as well as the
growth/damping rate, $|\gamma|$, becomes small compared to the real wave
frequency, $\omega_{r}$. This allows one to treat the wave phase-velocity
relation derived by the dominant real part of $D(\omega,\vec{k})$%
),\ separately from the instability driving derived by the small imaginary
part of $D(\omega,\vec{k})$.\ Second, equation~(\ref{D_reduced}) allows one to
expose all instability driving and loss mechanisms as separate linear terms.
This is convenient for the general instability analysis, although here we
restrict ourselves to the purely isothermal FBI.

Under condition of $|\gamma|\ll\omega_{r}$, if we also neglect the
corresponding first-order small terms in the RHS\ of equation~(\ref{D_reduced}%
) and substitute $\omega\approx\omega_{r}$ in all highest-order terms, we
obtain the equation for the real wave frequency, $\operatorname{Re}%
D(\omega,\vec{k})\approx D_{0}(\omega_{r},\vec{k})$. The solution of
$D_{0}(\omega_{r},\vec{k})=0$ for $\operatorname{Re}\omega=\omega_{r}$
provides the zeroth-order phase-velocity relations for the linear harmonic
waves, $\omega_{r}(\vec{k})$. In the next step, we add the small imaginary
parts and solve for the first-order equation with $i\gamma$ included in the
complex wave frequency. This gives
\begin{linenomath*}
\begin{equation}
\gamma\approx-\left.  \frac{\operatorname{Im}D(\omega,\vec{k})}{\partial
D_{0}(\omega,\vec{k})/\partial\omega}\right\vert _{\omega=\omega_{r}}.
\label{gamma_small}%
\end{equation}
\end{linenomath*}

The zeroth-order relation for the dominant real part of the wave frequency is
obtained by neglecting in the RHS\ of equation~(\ref{D_reduced}) all terms
proportional to $A_{j}$, except the ratios $A_{\mathrm{TB}}/A_{i}$ and
$A_{\mathrm{SE}}/A_{i}$. This yields
\begin{linenomath*}
\begin{equation}
D_{0}(\omega_{r},\vec{k})=1+(1+\kappa_{i}^{2})\left(  1+\frac{\vec{k}\cdot
\vec{U}}{\Omega_{e}}\right)  \Psi=0, \label{D_0}%
\end{equation}
\end{linenomath*}
where $\Omega_{e}$ is the common Doppler-shifted wave frequency for all
electrons, $\Omega_{e}=\Omega_{\mathrm{TB}}=\Omega_{\mathrm{SE}}\approx
\omega_{r}-\vec{k}\cdot\vec{V}_{0}$,
\begin{linenomath*}
\begin{equation}
\Psi\equiv\rho_{\mathrm{TB}}\psi_{\mathrm{TB}}+\rho_{\mathrm{SE}}%
\psi_{\mathrm{SE}},\qquad\psi_{j}\equiv\frac{1}{\kappa_{j}\kappa_{i}}\left(
1+\frac{\kappa_{j}^{2}k_{\parallel}^{2}}{k_{\perp}^{2}}\right)  , \label{Psi=}%
\end{equation}
\end{linenomath*}
and $\vec{U}$ is defined by equation~(\ref{UU}) with the use of the relation
$\Omega_{i}=\Omega_{e}+\vec{k}\cdot\vec{U}$. The solution of
equation~(\ref{D_0}) for $\Omega_{e}$ yields
\begin{linenomath*}
\begin{align}
&  \Omega_{e}(\vec{k})=-\ \frac{(1+\kappa_{i}^{2})(\vec{k}\cdot\vec{U})\Psi
}{1+(1+\kappa_{i}^{2})\Psi},\nonumber\\
&  \Omega_{i}(\vec{k})=\frac{\vec{k}\cdot\vec{U}}{1+(1+\kappa_{i}^{2})\Psi
},\label{Phase_velocity_relation}\\
&  \omega_{r}(\vec{k})=\frac{\vec{k}\cdot\lbrack\vec{V}_{0}+(1+\kappa_{i}%
^{2})\Psi\vec{V}_{i0}]}{1+(1+\kappa_{i}^{2})\Psi}.\nonumber
\end{align}
\end{linenomath*}
These expressions provide the phase-velocity relation, $\vec{V}_{\mathrm{ph}%
}=\omega_{r}(\vec{k})/\vec{k}$, in various frames of reference. For the low
ion magnetization, $\kappa_{i}^{2}\ll1$, $|\vec{V}_{i0}|\ll V_{0}$, $\vec
{U}_{i}\approx\vec{V}_{0}$, that usually takes place at E-region altitudes
below $115$\ km, we have much simpler relations:
\begin{linenomath*}
\begin{equation}
\Omega_{e}\approx-\ \frac{\Psi(\vec{k}\cdot\vec{V}_{0})}{1+\Psi},\qquad
\Omega_{i}\approx\omega_{r}\approx\frac{\vec{k}\cdot\vec{V}_{0}}{1+\Psi}.
\label{low_magnetized_ions_phase_velocity}%
\end{equation}
\end{linenomath*}
They become the conventional FBI expressions if the parameter $\Psi$ defined
in equation~(\ref{Psi=}) is replaced by the single-group parameter $\psi
_{e}=(\nu_{e}\nu_{i}/\omega_{ce}\omega_{ci})(1+\omega_{ce}^{2}k_{\parallel
}^{2}/\nu_{e}^{2}k_{\perp}^{2})$. Note also that in the long-wavelength limit
the phase-velocity relations given by (\ref{Phase_velocity_relation}%
)--(\ref{low_magnetized_ions_phase_velocity}) are common for all E-region instabilities.

Calculating the FBI growth rate requires more cumbersome algebra. Skipping
most of it, we obtain
\begin{linenomath*}
\begin{align*}
\operatorname{Im}D(\omega,\vec{k})  &  =\frac{1}{\nu_{i}\Omega_{i}}\left\{
\Omega_{i}^{2}-\left[  1+\frac{\rho_{\mathrm{TB}}\psi_{\mathrm{TB}}%
^{2}T_{\mathrm{TB}}+\rho_{\mathrm{SE}}\psi_{\mathrm{SE}}^{2}T_{\mathrm{SE}}%
}{\Psi^{2}T_{i}}\right]  k_{\perp}^{2}V_{Ti}^{2}\right\}  ,\\
\frac{\partial D_{0}(\omega_{r},\vec{k})}{\partial\omega_{r}}  &  =-\ \frac
{1}{\Psi(\vec{k}\cdot\vec{U}_{i})},
\end{align*}
\end{linenomath*}
where $\Omega_{i}$ is given by equation~(\ref{Phase_velocity_relation}) and
$V_{Ti}^{2}=T_{i}/m_{i}$. Then equation~(\ref{gamma_small}) yields
\begin{linenomath*}
\begin{align}
\gamma &  =\frac{\Psi}{\left[  1+\left(  1+\kappa_{i}^{2}\right)  \Psi\right]
\nu_{i}}\left\{  \left[  1-\kappa_{i}^{2}-\frac{\left(  1+\kappa_{i}%
^{2}\right)  ^{2}\nu_{i}^{2}}{\omega_{pi}^{2}}\right]  \Omega_{i}^{2}\right.
\nonumber\\
&  \left.  -\left(  1+\frac{\rho_{\mathrm{TB}}\psi_{\mathrm{TB}}%
^{2}T_{\mathrm{TB}}+\rho_{\mathrm{SE}}\psi_{\mathrm{SE}}^{2}T_{\mathrm{SE}}%
}{\Psi^{2}T_{i}}\right)  k_{\perp}^{2}V_{Ti}^{2}\right\}  ,
\label{gamma_fin_2e_fluid}%
\end{align}
\end{linenomath*}
where $\omega_{pi}=[n_{0}e^{2}/(\epsilon_{0}m_{i})]^{1/2}$ is the ion plasma
frequency. Equating $\gamma=0$, we obtain the expression for $\Omega_{i}%
(\vec{k})$ and, through equations~(\ref{UU}) and
(\ref{Phase_velocity_relation}), the threshold values of the FBI driving field, $E_{\mathrm{Thr}%
}(\vec{k})$.

We will not analyze here expression~(\ref{gamma_fin_2e_fluid}) because the three-fluid model is too
oversimplified and cannot provide accurate quantitative description of the FBI
in the presence of the electron precipitation. The main point of this tentative
analysis was to demonstrate the unexpected complexity of the parameter
dependence. The contribution of the partial parameters $\psi_{\mathrm{TB}}$
and $\psi_{\mathrm{TB}}$ into the total parameter $\Psi$, as described by
equation~(\ref{Psi=}), is natural and well-expected generalization. What was
not expected though is the explicit involvement of these partial parameters
into the weighting factors for $T_{\mathrm{TB}}$ and $T_{\mathrm{SE}}$ within
the FBI diffusion loss term $\propto k_{\perp}^{2}V_{Ti}^{2}$.
If there were no additional weighting factors
$\psi_{\mathrm{TB}}^{2}$ and $\psi_{\mathrm{SE}}^{2}$ then the three-fluid model
expression for the growth/damping rate $\gamma$ would correspond to the
naively suggested replacement of the electron temperature with the effective
temperature $T_{\mathrm{tot}}$ defined by equation~(\ref{T_tot}). However, the
additional multipliers $\psi_{\mathrm{TB}}^{2}$ and $\psi_{\mathrm{SE}}^{2}$,
each proportional to $\nu_{\mathrm{TB}}^{2}$ and $\nu_{\mathrm{SE}}^{2}$,
respectively, makes the contribution of each electron group into the FBI\ loss
term much less obvious with potentially significant quantitative consequences.

Since the $e$-$n$ collision frequency has a gradual dependence on the
individual electron velocity, this dependence cannot be accurately reduced to
just two different constant values $\nu_{\mathrm{TB}}$ and $\nu_{\mathrm{SE}}%
$, as we have done in this tentative analysis. The only proper and accurate way to
quantitatively treat electron collisions for the two-component electron
distribution with non-Maxwellian high-energy tail is by employing the rigorous
kinetic theory. This will be done in the following section.

\subsection{Kinetic Analysis of the FBI Onset for the General Electron
Distribution Function\label{Kinetic analysis of the FB instability onset for the general electron distribution function}}

In this section, we develop a kinetic linear theory of the FBI for a system with general
non-Maxwellian electron distributions. Our hybrid theoretical
approach combines the fully kinetic description of electrons with the
fluid-model description of ions.

As above, we will restrict our treatment to the E-region altitudes where
the electrons are highly magnetized, $\omega_{ec}\gg\nu_{en}$, while ions are at
least partially unmagnetized, $\omega_{ic}\lesssim\nu_{in}$. These conditions
typically hold within the core of the high-latitude E-region ionosphere between
90 and 120~km. Under these magnetization conditions, electrons
are essentially $\vec{E}_{0}\times\vec{B}$ drift, while ions mostly move with the
dominant neutral component. The linear instability onset is described by
analyzing small harmonic wave perturbations of the plasma particle motion and the
coupled electrostatic potential.

Before proceeding with the fluid-model ion description, we introduce
dimensionless variables and parameters:
\begin{linenomath*}
\begin{equation}
\eta_{\omega,\vec{k}}\equiv\frac{\delta n_{\omega,\vec{k}}}{n_{0}},\qquad
\phi_{\omega,\vec{k}}\equiv\frac{e\Phi_{\omega,\vec{k}}}{T_{e0}},\qquad
\beta_{T}\equiv\frac{T_{i0}}{T_{e0}}, \label{dimensional}%
\end{equation}
\end{linenomath*}
where $T_{e0}$ is an effective electron temperature, while the entire ion
population is assumed to be Maxwellian with the constant temperature $T_{i0}$.
For the general non-Maxwellian electron distribution, we will not specify the
parameter $T_{e0}$. We have introduced it here as a convenient normalization
constant, but the final expressions will not depend on $T_{e0}$.

For isothermal ions, the fluid model equations (\ref{continuity_s}) and
(\ref{momentum_s}) yield the following relation between $\phi_{\omega,\vec{k}%
}$ and $\eta_{\omega,\vec{k}}$:
\begin{linenomath*}
\begin{equation}
\phi_{\omega,\vec{k}}\approx\beta_{T}\left\{  \frac{\left[  \left(
1-\kappa_{i}^{2}\right)  \Omega_{\omega,k}+i(1+\kappa_{i}^{2})\nu_{in}\right]
\Omega_{\omega,\vec{k}}}{k^{2}V_{Ti}^{2}}-1\right\}  \eta_{\omega,\vec{k}},
\label{from_ions}
\end{equation}
\end{linenomath*}
where $\Omega_{\omega,\vec{k}}\equiv\omega-\vec{k}\cdot\vec{V}_{i0}$ is the
Doppler-shifted wave frequency in the ion-flow frame of reference, moving
relative the neutrals with the mean flow velocity,
\begin{linenomath*}
\begin{equation}
\vec{V}_{i0}=\left.  \left(  \frac{e\vec{E}_{0}}{m_{i}\nu_{in}}+\kappa_{i}%
^{2}\vec{V}_{0}\right)  \right/  (1+\kappa_{i}^{2}), \label{V_i0}%
\end{equation}
\end{linenomath*}
and $V_{Ti}=(T_{i0}/m_{i})^{1/2}$ is the ion
thermal velocity. Recall that $\vec{V}_{0}=\vec{E}_{0}\times\vec{B}/B^{2}$ is
the $\vec{E}_{0}\times\vec{B}$-drift velocity and $\kappa_{i}=\omega_{ic}%
/\nu_{in}$ is the ion magnetization parameter. The full expression for
$\vec{V}_{i0}$ is of importance only for altitudes above 115~km where
$\kappa_{i}\gtrsim1$. At altitudes well below 115~km, the ions are essentially
unmagnetized, $\kappa_{i}\ll1$, $\nu_{in}\gg\omega_{ic}$, so that the mean
ion-flow speed is negligibly small compared the mean speed of the highly
magnetized electrons $V_{0}=|\vec{V}_{0}|=eE_{0}/(m_{i}\omega_{ic})$.

Here we generalize the kinetic description of electrons in \citeA{Dimant:Kinetic95a} by assuming
arbitrary ion magnetization and, more importantly, by assuming non-Maxwellian
velocity distribution of the background electrons.

If we suppose that collisional angular scattering of electrons is much faster than
the corresponding energy changes then the electron velocity distribution
consists mostly of two different parts,
\begin{linenomath*}
\begin{equation}
f_{e}(\vec{V},\vec{r})\approx F_{0}(V,\vec{r})+\frac{\vec{f}_{1}(V,\vec
{r})\cdot\vec{V}}{V},\qquad|\vec{f}_{1}(V,\vec{r})|\ll F_{0}(V,\vec{r}),
\label{simple}%
\end{equation}
\end{linenomath*}
where $\vec{V}$ is the electron velocity and $V$ is the
corresponding speed.
The function $F_{0}(V,\vec{r})$ is the
dominant omnidirectional, i.e., isotropic, part of $f_{e}(\vec{V},\vec{r})$, while
$\vec{f}_{1}(V,\vec{r})$ is a small directional part. The isotropic part
$F_{0}(V,\vec{r})$ is responsible for the integral scalar quantities like the
total electron density, pressure,
\begin{linenomath*}
\begin{equation}
n_{e}(\vec{r})\approx4\pi\int_{0}^{\infty}F_{0}(V,\vec{r})V^{2}dV,\qquad
P_{e}(\vec{r})\approx\frac{4\pi m_{e}}{3}\int_{0}^{\infty}F_{0}(V,\vec
{r})V^{4}dV, \label{n_e,P_e}%
\end{equation}
\end{linenomath*}
and temperature, $T_{\mathrm{tot}}(\vec{r})=P_{e}(\vec{r})/n_{e}(\vec{r})$,
while the small directional part is responsible for various fluxes and
currents, like the total particle number flux,
\begin{linenomath*}
\begin{equation}
\vec{\Gamma}(\vec{r})\approx\frac{4\pi}{3}\int_{0}^{\infty}\vec{f}_{1}%
(V,\vec{r})V^{3}dV \label{Gamma(r)}%
\end{equation}
\end{linenomath*}
and the corresponding energy flux. Equation (\ref{simple}) represents the two
highest-order terms of the Legendre polynomial series
\cite{Shkarofsky:Particle66,Gurevich:Nonlinear78,Khazanov:Kinetic11}. No higher-order
angular dependencies of $f_{e}(\vec{V},\vec{r})$, responsible for the
anisotropic pressure, viscosity tensors, etc., are included in approximate
equation (\ref{simple}).

Under conditions of strong isotropization of the electron distribution
function, the general kinetic equation
reduces to a set of two coupled integro-differential equations
\cite[equations (11) and (12)]{Dimant:Kinetic95a}, whose further
simplification results in $\vec{f}_{1}(V,\vec{r})$ explicitly expressed in
terms of $F_{0}(V,\vec{r})$ \cite[equations
(13) and (14)]{Dimant:Kinetic95a}. This allows one to obtain a closed partial differential
equation for $F_{0}(V,\vec{r})$,
\begin{linenomath*}
\begin{equation}
\left(  \frac{\partial}{\partial t}+\vec{V}_{\mathrm{dr}}\cdot\nabla+\hat
{R}\right)  F_{0}(V,\vec{r})=0, \label{For_F_0}%
\end{equation}
\end{linenomath*}
where the total $\vec{E}\times\vec{B}$-drift velocity is given by
\begin{linenomath*}
\begin{equation}
\vec{V}_{\mathrm{dr}}\equiv\vec{V}_{0}+\frac{e}{m_{e}\omega_{ec}}\ \hat
{b}\times\nabla\Phi, \label{V_dr}%
\end{equation}
\end{linenomath*}
and the differential operator $\hat{R}$ is given by
\begin{linenomath*}
\begin{align}
&  \hat{R}\equiv-\ \frac{1}{3V^{2}}\left(  \mathbf{\hat{K}}_{\perp}\cdot
\frac{V^{2}\nu_{en}(V)}{\omega_{ec}^{2}}\ \mathbf{\hat{K}}_{\perp}+\hat
{K}_{\parallel}\ \frac{V^{2}}{\nu_{en}(V)}\ \hat{K}_{\parallel}\right)
,\nonumber\\
&  \mathbf{\hat{K}}_{\perp}\equiv V\nabla_{\perp}-\frac{e\vec{E}_{\perp}%
}{m_{e}}\frac{\partial}{\partial V},\qquad\hat{K}_{\parallel}\equiv
V\ \frac{\partial}{\partial z}-\frac{eE_{\parallel}}{m_{e}}\frac{\partial
}{\partial V}, \label{R,K}%
\end{align}
\end{linenomath*}
where $\partial/\partial z$ is the derivative in the $\vec{B}$ direction;
$E_{\parallel}$ and $\vec{E}_{\perp}$ are the parallel and perpendicular to
$\vec{B}$ components of the total electrostatic field, $\vec{E}=\vec{E}%
_{0}-\nabla\Phi$; and $\hat{b}=\vec{B}/b$ is the unit vector in the $\vec{B}$
direction. Equations~(\ref{For_F_0})--(\ref{R,K}) differ from \citeA[equations~(16)--(23)]{Dimant:Kinetic95a}
by some notations and, most importantly, by neglecting
here the $e$-$e$ collisions and the terms describing the thermal exchange between
electrons and neutrals through $e$-$n$ collisions. The former is important for
sufficiently dense and low-energy particles, while the latter is crucial for
the electron thermal instability (ETI) \cite[and references therein]{Dimant:Kinetic95b,Dimant:Kinetic95,
Dimant:Physical97,Oppenheim:Newly20}.
Bearing in mind the pure FBI, we disregard here any thermal-instability effects.

Equation (\ref{For_F_0}) holds for the entire isotropic part of the electron
distribution function, $F_{0}(V,\vec{r})$, which includes the spatially
homogeneous background distribution, $f_{0}(V)$, and all linear wave
perturbations, $f_{\omega,\vec{k}}(V)$. Linearizing this equation for a given
wave harmonic, after some algebra we arrive at equation~(38) from
\citeA{Dimant:Kinetic95a}:
\begin{linenomath*}
\begin{equation}
\left(i\Delta_{\omega,\vec{k}}+\hat{D}_{\omega,\vec{k}}\right)
f_{\omega,\vec{k}}(V)=\left(  \hat{B}_{\omega,\vec{k}}F_{0}(V)\right)
\phi_{\omega,\vec{k}}, \label{urka}%
\end{equation}
\end{linenomath*}
where various differential operators acting on both $f_{\omega,\vec{k}}$ and
$F_{0}$ are defined by
\begin{linenomath*}
\begin{subequations}
\label{DB}%
\begin{align}
\hat{D}_{\omega,\vec{k}}  &  \equiv-\ \frac{1}{3V^{2}}\left(  \mathbf{\hat{K}%
}_{\perp}^{(0)}\cdot\frac{V^{2}\nu_{en}(V)}{\omega_{ec}^{2}}\ \mathbf{\hat{K}%
}_{\perp}^{(0)}+\hat{K}_{\parallel}^{(0)}\ \frac{V^{2}}{\nu_{en}(V)}\ \hat
{K}_{\parallel}^{(0)}\right) \label{D}\\
\Delta_{\omega,\vec{k}}  &  \equiv\vec{k}\cdot\vec{V}_{0}-\omega
,\qquad\mathbf{\hat{K}}_{\perp}^{(0)}=i\vec{k}_{\perp}V-\frac{e\vec{E}_{0}%
}{m_{e}}\frac{d}{dV},\qquad\hat{K}_{\parallel}^{(0)}=ik_{\parallel
}V,\label{Delta,K^(0)}\\
\hat{B}_{\omega,\vec{k}}  &  \equiv-\ \frac{T_{e0}}{3m_{e}}\left[  \left(
\frac{k_{\perp}^{2}\nu_{en}(V)}{\omega_{ec}^{2}}+\ \frac{k_{\parallel}^{2}%
}{\nu_{en}(V)}\right)  V\ \frac{d}{dV}\right. \nonumber\\
&  +\left.  2i\ \frac{e\vec{k}_{\perp}\cdot\vec{E}_{0}}{m_{e}\omega_{ec}%
^{2}V^{2}}\ \frac{d}{dV}\left(  V^{2}\nu_{en}(V)\frac{d}{dV}\right)  \right]
. \label{B_omega,k}%
\end{align}
\end{subequations}
\end{linenomath*}
Equation~(\ref{urka}) implies an arbitrary background distribution function
$F_{0}(V)$ that provides convergence of any integrals, like those in
equation (\ref{n_e,P_e}). In our case, $F_{0}(V)$ includes both the low-energy
Maxwellian bulk distribution and the high-energy superthermal tail, We will
specify these components later, but now will proceed with arbitrary $F_{0}%
(V)$. Note that the definition $\hat{B}_{\omega,\vec{k}}$ includes the
normalization constant $T_{e0}$ in the numerator, while the wave potential
$\phi_{\omega,\vec{k}}$, defined in equation~(\ref{dimensional}), contains
$T_{e0}$ in the denominator, so that the RHS\ of (\ref{urka}) is actually
$T_{e0}$-independent.

In accord with the above discussion, we drop all terms proportional to $\vec{k}_{\perp}%
\cdot\vec{E}_{0}$ because these terms describe frictional heating and will
eventually lead to the ETI. The pure
FBI is described by the remaining terms, like the first term in the RHS of
equation~(\ref{B_omega,k}), which is proportional to $Vd/dV$.

Now we obtain the second relation between $\eta_{\omega,\vec{k}}$\ and
$\phi_{\omega,\vec{k}}$, analogous to equation (\ref{from_ions}). Before
proceeding, we specify the main conditions for the vast majority of the
FBI-driven waves. These low-frequency and long-wavelength waves usually
satisfy
\begin{linenomath*}
\begin{equation}
k_{\parallel}\ll k_{\perp}\approx k,\qquad\gamma\ll\omega,~kV_{0},\ll\nu
_{in},\qquad k\lambda_{D}\ll1, \label{k_conditions}%
\end{equation}
\end{linenomath*}
so that the following inequality holds,
\begin{linenomath*}
\begin{equation}
|\hat{D}_{\omega,\vec{k}}|\ll|\Delta_{\omega,\vec{k}}|. \label{D,B<<Delta}%
\end{equation}
\end{linenomath*}
These symbolic relation means that the operators $\hat{D}_{\omega,\vec{k}}$
and $\Delta_{\omega,\vec{k}}$\ apply to $f_{\omega,\vec{k}}(V)$ and the
results are compared by the absolute value. The main point of
equation~(\ref{D,B<<Delta}) is that one can apply to equation~(\ref{urka}) a
formal Taylor expansion with respect to the small ratio $|\hat{D}_{\omega,\vec{k}%
}|/|\Delta_{\omega,\vec{k}}|$. This leads to
\begin{linenomath*}
\begin{equation}
f_{\omega,\vec{k}}(V)\approx\left(  \frac{\hat{B}_{\omega,\vec{k}}F_{0}%
(V)}{i\Delta_{\omega,\vec{k}}}+\frac{\hat{D}_{\omega,\vec{k}}\hat{B}%
_{\omega,\vec{k}}F_{0}(V)}{\Delta_{\omega,\vec{k}}^{2}}\right)  \phi
_{\omega,\vec{k}}, \label{fvsPhi}%
\end{equation}
\end{linenomath*}
where the order of the two differential operators $\hat{D}_{\omega,\vec{k}}$
and $\hat{B}_{\omega,\vec{k}}$ matters. This formal expansion procedure is
equivalent to a regular perturbation technique when one initially neglects in
equation~(\ref{urka}) the term $\propto\hat{D}_{\omega,\vec{k}}$ and then
finds the first-order correction. Proceeding from $f_{\omega,\vec{k}}$ to
$\eta_{\omega,\vec{k}}$ through the relation
\begin{linenomath*}
\[
\eta_{\omega,\vec{k}}=\frac{4\pi\int_{0}^{\infty}f_{\omega,\vec{k}}(V)V^{2}%
dV}{n_{0}}=\frac{\int_{0}^{\infty}f_{\omega,\vec{k}}(V)V^{2}dV}{\int%
_{0}^{\infty}F_{0}(V)V^{2}dV},
\]
\end{linenomath*}
following from equations~(\ref{dimensional}) and (\ref{n_e,P_e}), we obtain
\begin{linenomath*}
\begin{equation}
\eta_{\omega,\vec{k}}\approx\frac{\langle\hat{B}_{\omega,\vec{k}}\rangle
}{i\Delta_{\omega,\vec{k}}}\left(  1+\frac{i\langle\hat{D}_{\omega,\vec{k}%
}\hat{B}_{\omega,\vec{k}}\rangle}{\Delta_{\omega,\vec{k}}\langle\hat
{B}_{\omega,\vec{k}}\rangle}\right)  \phi_{\omega,\vec{k}}, \label{promeha}%
\end{equation}
\end{linenomath*}
where the speed averaging of any operator or function $\hat{A}$ is defined as
\begin{linenomath*}
\begin{equation}
\left\langle \hat{A}\right\rangle \equiv\frac{4\pi}{n_{0}}\int_{0}^{\infty
}\left(  \hat{A}F_{0}(V)\right)  V^{2}dV=\frac{\int_{0}^{\infty}\left(
\hat{A}F_{0}(V)\right)  V^{2}dV}{\int_{0}^{\infty}F_{0}(V)V^{2}dV}.
\label{<A>}%
\end{equation}
\end{linenomath*}
Expressing $\phi_{\omega,\vec{k}}$ again to the first-order accuracy with
respect to the small parameter $|\langle\hat{D}_{\omega,\vec{k}}\hat
{B}_{\omega,\vec{k}}\rangle|/|\Delta_{\omega,\vec{k}}\langle\hat{B}%
_{\omega,\vec{k}}\rangle|$ we obtain equation~(52) from
\citeA{Dimant:Kinetic95a}:
\begin{linenomath*}
\begin{equation}
\phi_{\omega,\vec{k}}\approx\left(  \frac{i\Delta_{\omega,\vec{k}}}%
{\langle\hat{B}_{\omega,\vec{k}}\rangle}+\frac{\langle\hat{D}_{\omega,\vec{k}%
}\hat{B}_{\omega,\vec{k}}\rangle}{\langle\hat{B}_{\omega,\vec{k}}\rangle^{2}%
}\right)  \eta_{\omega,\vec{k}}. \label{soot}%
\end{equation}
\end{linenomath*}

Combining equations~(\ref{from_ions}) and (\ref{soot}), we obtain the FBI
dispersion relation:
\begin{linenomath*}
\begin{equation}
\frac{\Omega_{\omega,\vec{k}}\left[  \left(  1-\kappa_{i}^{2}\right)
\Omega_{\omega,k}+i\nu_{in}(1+\kappa_{i}^{2})\right]  }{k^{2}V_{Ti}^{2}%
}=1+\frac{1}{\beta_{T}}\left(  \frac{\langle\hat{D}_{\omega,\vec{k}}\hat
{B}_{\omega,\vec{k}}\rangle}{\langle\hat{B}_{\omega,\vec{k}}\rangle^{2}}%
+\frac{i\Delta_{\omega,\vec{k}}}{\langle\hat{B}_{\omega,\vec{k}}\rangle
}\right)  , \label{disperga}%
\end{equation}
\end{linenomath*}
which generalizes equation~(58) from \citeA{Dimant:Kinetic95a} for
general $F_{0}(V)$ and arbitrary ion magnetization.

The differential operators $\hat{D}_{\omega,\vec{k}}$ and $\hat{B}%
_{\omega,\vec{k}}$ are defined by equation~(\ref{DB}). The terms proportional
to $\vec{E}_{0}$ are crucial for the ETI, but for the FBI they play no role,
so that we can reduce these operators to simpler expressions,
\begin{linenomath*}
\begin{equation}
\hat{D}_{\omega,\vec{k}}\Rightarrow\frac{k_{\perp}^{2}V^{2}m_{e}\psi_{e}%
(V)}{3m_{i}\nu_{in}},\qquad\hat{B}_{\omega,\vec{k}}\Rightarrow-\ \frac
{k_{\perp}^{2}T_{e0}}{3m_{i}\nu_{in}}\ V\psi_{e}(V)\ \frac{d}{dV},
\label{DB=>}%
\end{equation}
\end{linenomath*}
where
\begin{linenomath*}
\begin{equation}
\psi_{e}(V)\equiv\frac{\nu_{en}(V)\nu_{in}}{\omega_{ec}\omega_{ic}}\left(
1+\frac{k_{\parallel}^{2}\omega_{ec}^{2}}{k_{\perp}^{2}\nu_{en}^{2}%
(V)}\right)  \label{psi_e(V)}%
\end{equation}
\end{linenomath*}
is the kinetic analog of the standard fluid-model parameter $\psi$ defined by
equation~(\ref{psi_standard}). The reduced expression for $\hat{D}%
_{\omega,\vec{k}}$ fully agrees with equation~(53) from
\citeA{Dimant:Kinetic95a} after neglecting in that equation the term
$\propto\zeta_{\omega,\vec{k}}$. At the same time, the operator $\hat
{B}_{\omega,\vec{k}}$ can be reduced to \citeA[equation~(54)]{Dimant:Kinetic95a},
only for Maxwellian $F_{0}(V)$. For general $F_{0}(V)$, in accord with
equation~(\ref{<A>}), we obtain after integration by parts:
\begin{linenomath*}
\begin{align}
\langle\hat{B}_{\omega,\vec{k}}\rangle &  =\frac{4\pi T_{e0}k_{\perp}^{2}%
}{3n_{0}m_{i}\nu_{in}}\int_{0}^{\infty}F_{0}(V)\ \frac{d\left(  V^{3}\psi
_{e}\right)  }{dV}\ dV,\nonumber\\
\left\langle \hat{D}_{\omega,\vec{k}}\hat{B}_{\omega,\vec{k}}\right\rangle  &
=\frac{4\pi T_{e0}m_{e}k_{\perp}^{4}}{9n_{0}m_{i}^{2}\nu_{in}^{2}}\int%
_{0}^{\infty}F_{0}(V)\ \frac{d\left(  V^{5}\psi_{e}^{2}\right)  }{dV}\ dV.
\label{BDB}%
\end{align}
\end{linenomath*}

Under conditions of equation~(\ref{k_conditions}), in both sides of dispersion equation
(\ref{disperga}) the imaginary parts dominate. This allows us to easily
separate the wave phase-velocity relation, $\omega_{r}(\vec{k})$, from the
wave growth/damping relation, $\gamma(\vec{k})$.

The wave phase-velocity relation is obtained to the zeroth-order accuracy,
after neglecting the small real parts, as well as small $\gamma$ in
$\omega=\omega_{r}+i\gamma$. This yields:
\begin{linenomath*}
\begin{equation}
\omega_{r}=\frac{\vec{k}\cdot\lbrack\vec{V}_{0}+(1+\kappa_{i}^{2})\tilde{\psi
}\vec{V}_{i0}]}{1+(1+\kappa_{i}^{2})\tilde{\psi}}, \label{omega_r}%
\end{equation}
\end{linenomath*}
where the constant parameter
\begin{linenomath*}
\begin{equation}
\tilde{\psi}=\frac{\beta_{T}\nu_{in}\langle\hat{B}_{\omega,\vec{k}}\rangle
}{k^{2}V_{Ti}^{2}}=\frac{4\pi}{3n_{0}}\int_{0}^{\infty}F_{0}(V)\ \frac
{d\left(  V^{3}\psi_{e}\right)  }{dV}\ dV \label{psi_new}%
\end{equation}
\end{linenomath*}
unlike $\psi_{e}(V)$, generalizes the conventional parameter $\psi$ for the
entire electron population. Equations~(\ref{V_i0}) and (\ref{omega_r}) yield
the real part of $\Omega_{\omega,\vec{k}}$:
\begin{linenomath*}
\begin{subequations}
\label{Omega_r,U=}%
\begin{align}
&  (\Omega_{\omega,\vec{k}})_{r}=\omega_{r}-\vec{k}\cdot\vec{V}_{i0}%
\approx\frac{\vec{k}\cdot\vec{U}}{1+(1+\kappa_{i}^{2})\tilde{\psi}%
},\label{Omega_r}\\
&  \vec{U}\equiv\vec{V}_{0}-\vec{V}_{i0}\approx\left.  \left(  \vec{V}%
_{0}-\frac{e\vec{E}_{0}}{m_{i}\nu_{in}}\right)  \right/  (1+\kappa_{i}^{2}).
\label{U=}%
\end{align}
\end{subequations}
\end{linenomath*}

To the first-order accuracy, equation~(\ref{disperga}) yields:
\begin{linenomath*}
\begin{equation}
\gamma\approx\frac{\tilde{\psi}\left[  (1-\kappa_{i}^{2})(\Omega_{\omega
,\vec{k}})_{r}^{2}-k^{2}\tilde{C}_{s}^{2}\right]  }{(1+\tilde{\psi})\nu_{in}},
\label{gamma}%
\end{equation}
\end{linenomath*}
where $\tilde{C}_{s}$ is a modified ion-acoustic speed,
\begin{linenomath*}
\begin{equation}
\tilde{C}_{s}^{2}=V_{Ti}^{2}+\frac{\langle\hat{D}_{\omega,\vec{k}}\hat
{B}_{\omega,\vec{k}}\rangle V_{Ti}^{2}}{\beta_{T}\langle\hat{B}_{\omega
,\vec{k}}\rangle^{2}}. \label{C_s^2_kinetic}%
\end{equation}
\end{linenomath*}
Using equation~(\ref{BDB}), the second term in the RHS of
equation~(\ref{C_s^2_kinetic}) can be written as
\begin{linenomath*}
\begin{align}
&  \frac{\langle\hat{D}_{\omega,\vec{k}}\hat{B}_{\omega,\vec{k}}\rangle
V_{Ti}^{2}}{\beta_{T}\langle\hat{B}_{\omega,\vec{k}}\rangle^{2}}=\frac
{m_{e}n_{0}}{4\pi m_{i}}\frac{\int_{0}^{\infty}F_{0}(V)\ [d\left(  V^{5}%
\psi_{e}^{2}\right)  /dV]\ dV}{\left(  \int_{0}^{\infty}F_{0}(V)\ [d\left(
V^{3}\psi_{e}\right)  /dV]\ dV\right)  ^{2}}\nonumber\\
&  =\frac{4\pi}{9n_{0}\tilde{\psi}^{2}}\frac{m_{e}}{m_{i}}\int_{0}^{\infty
}F_{0}(V)\ \frac{d\left(  V^{5}\psi_{e}^{2}\right)  }{dV}\ dV.
\label{second_term}%
\end{align}
\end{linenomath*}

Equations~(\ref{omega_r}) and (\ref{C_s^2_kinetic}) totally agree with
equations~(5) and (6) from \citeA{Dimant:Model03} after replacing there the standard
parameters $\psi$ and $C_{s}^{2}$ with $\tilde{\psi}$ and $\tilde{C}_{s}^{2}$,
respectively. Notice that if the ion magnetization is sufficiently high,
$\kappa_{i}>1$, then the FBI driving mechanism, described in
equation~(\ref{gamma}) by the term $(1-\kappa_{i}^{2})(\Omega_{\omega,\vec{k}})_{r}^{2}$,
becomes stabilizing, as discussed in detail in \citeA{Dimant:Ion04}. This
happens above the magnetization boundary, $\kappa_{i}=1$, which at the
high-latitude ionosphere is located about 120~km of altitude \cite<e.g.,>[Fig.~5]{Dimant:Ion04}.
In this paper, we will restrict our analysis to lower
E-region altitudes where $\kappa_{i}<1$.

We can rewrite the expression for the modified ion-acoustic speed, $\tilde
{C}_{s}$, in a more traditional way as
\begin{linenomath*}
\begin{equation}
\tilde{C}_{s}=\left(  \frac{T_{i}+T_{\mathrm{eff}}}{m_{i}}\right)
^{1/2},\qquad T_{\mathrm{eff}}=\frac{4\pi m_{e}}{9n_{0}\tilde{\psi}^{2}}%
\int_{0}^{\infty}F_{0}(V)\ \frac{d\left(  V^{5}\psi_{e}^{2}\right)  }{dV}\ dV.
\label{Cs^tilde^2,gamma^tilde}%
\end{equation}
\end{linenomath*}
Emphasize that $\tilde{C}_{s}$ is not the actual ion-acoustic speed because in
the highly dissipative lower ionosphere no ion-acoustic wave can survive for a
time duration $\gtrsim\nu_{in}^{-1}$. For ion-acoustic waves, the
collisional damping is even more detrimental than the collisionless ion Landau
damping at much higher ionospheric altitudes (if there $T_{e}\sim T_{i}$). In
the highly dissipative E-region ionosphere, the analogs of the
ion-acoustic-like waves are precisely the compression/decompression waves
driven by the FBI and other plasma instabilities. These waves, however, can
survive for a time duration much longer than $\nu_{in}^{-1}$ only because
they are sustained by the external DC electric field, $\vec{E}_{0}\perp\vec
{B}$.

For constant $\nu_{en}$ (and hence for constant $\psi_{e}$), the above
expressions reduce to the fluid-model FBI\ wave phase velocity and
growth/damping rate relations. Indeed, in this case equation~(\ref{psi_new})
yields $\tilde{\psi}=\psi_{e}=\psi$, so that equation~(\ref{omega_r}) reduces
to the fluid-model phase-velocity relation, see, e.g., equation~(5) from
\citeA{Dimant:Model03}, even for arbitrary background electron distribution function $F_{0}(V)$.
For constant $\psi_{e}$, equation~(\ref{gamma}) reduces to fluid-model
equation~(6) from \citeA{Dimant:Model03} for isothermal ions and adiabatic electrons,
\begin{linenomath*}
\begin{equation}
\gamma=\frac{\psi\lbrack(1-\kappa_{i}^{2})(\Omega_{\omega,\vec{k}})_{r}%
^{2}-k^{2}C_{s}^{2}]}{(1+\psi)\nu_{in}},\qquad C_{s}^{2}=\frac{T_{i}%
+(5/3)T_{\mathrm{tot}}}{m_{i}}, \label{gamma_standard}%
\end{equation}
\end{linenomath*}
where $T_{\mathrm{tot}}$ is defined by equation~(\ref{T_tot}). In reality,
however, the kinetic quantity $\nu_{en}$ is strongly velocity-dependent, so that the
exact form of the omnidirectional function $F_{0}(V)$ does really matter.

The fact that for constant $\nu_{en}$ the electron temperature term in
$C_{s}^{2}$ includes the single-atom adiabaticity coefficient $5/3$ is
associated with the fact that we have neglected here the frictional heating
and the corresponding collisional cooling of electrons. This approximation
works for waves having sufficiently high wave frequencies, $\omega,$
$kV_{0}\gg\delta_{en}\nu_{en}$, while still satisfying the low-frequency,
long-wavelength conditions imposed by equation~(\ref{k_conditions}). Here
$\delta_{en}\simeq(2$--$4)\times10^{-3}$ is the mean relative fraction of
collisional losses of the electron energy during one $e$-$n$ collision
\cite{Gurevich:Nonlinear78,Dimant:Kinetic95a}. In the opposite
limit of very low-frequency, long-wavelength waves, $\omega,$ $kV_{0}\ll
\delta_{en}\nu_{en}$, the electron thermal behavior is mostly determined by
the heating/cooling balance, so that the factor $5/3$ disappears and the
destabilizing ETI mechanism for the optimal $\vec{k}$ directions becomes
efficient \cite{Dimant:Kinetic95b,Dimant:Kinetic95,Dimant:Physical97}.
The net result of this change is that
the minimum threshold field is reached for longer-wavelength waves than for
those prone to the pure FBI\ excitation.

The linear instability develops if the DC electric field exceeds the threshold
field determined by $\gamma=0$. According to equations~(\ref{Omega_r,U=}) and
(\ref{gamma}), this yields the threshold parameters
\begin{linenomath*}
\[
\frac{E_{\mathrm{Thr}}}{B}=V_{\mathrm{Thr}}=\frac{(1+\kappa_{i}^{2})\left[
1+(1+\kappa_{i}^{2})\tilde{\psi}\right]  }{\left(  \cos\theta-\kappa_{i}%
\sin\theta\right)  \sqrt{1-\kappa_{i}^{2}}}\ \tilde{C}_{s},
\]
\end{linenomath*}
where $\theta$ is the angle between the wavevector $\vec{k}$ and the $\vec
{E}_{0}\times\vec{B}$-drift direction (the \textquotedblleft
flow\textquotedblright\ angle).

Crucial for the onset of the FBI is the
minimum threshold field at a given location. The driving field and the
corresponding $\vec{E}_{0}\times\vec{B}$-drift speed reach their minimal
values at the optimal direction of the wavevector, $\vec{k}\parallel\vec{U}$,
corresponding to $\theta=-\arctan\kappa_{i}$ and $k_{\parallel}=0$:
\begin{linenomath*}
\begin{align}
&  \frac{(E_{\mathrm{Thr}})_{\min}}{B}=(V_{\mathrm{Thr}})_{\min}=\sqrt
{\frac{1+\kappa_{i}^{2}}{1-\kappa_{i}^{2}}}\left[  1+(1+\kappa_{i}^{2}%
)\tilde{\psi}\right]  \tilde{C}_{s}\nonumber\\
&  =\sqrt{\frac{1+\kappa_{i}^{2}}{1-\kappa_{i}^{2}}}\left[  1+(1+\kappa
_{i}^{2})\frac{I_{1}}{3I_{0}}\right]  \left(  \frac{T_{i}}{m_{i}}+\frac
{m_{e}I_{0}I_{2}}{m_{i}I_{1}^{2}}\right)  ^{1/2}, \label{E_thr_min}%
\end{align}
\end{linenomath*}
where in the last equality we expressed $\tilde{\psi}$ (for $k_{\parallel}=0$)
and $\tilde{C}_{s}$ in terms of the following integral parameters:
\begin{linenomath*}
\begin{subequations}
\label{I_0,1,2}%
\begin{align}
I_{0}  &  =\frac{n_{0}}{4\pi}=\int_{0}^{\infty}F_{0}V^{2}dV=\frac{\sqrt{2}%
}{m_{e}^{3/2}}\int_{0}^{\infty}F_{0}\sqrt{\mathcal{E}}d\mathcal{E}%
,\label{I_0}\\
I_{1}  &  =\int_{0}^{\infty}F_{0}\ \frac{d\left(  V^{3}\psi_{e}\right)  }%
{dV}\ dV=\left(  \frac{2}{m_{e}}\right)  ^{3/2}\int_{0}^{\infty}F_{0}%
\ \frac{d\left(  \mathcal{E}^{3/2}\psi_{e}\right)  }{d\mathcal{E}%
}\ d\mathcal{E},\label{I_1}\\
I_{2}  &  =\int_{0}^{\infty}F_{0}\ \frac{d\left(  V^{5}\psi_{e}^{2}\right)
}{dV}\ dV=\left(  \frac{2}{m_{e}}\right)  ^{5/2}\int_{0}^{\infty}F_{0}%
\ \frac{d\left(  \mathcal{E}^{5/2}\psi_{e}^{2}\right)  }{d\mathcal{E}%
}\ d\mathcal{E}. \label{I_2}%
\end{align}
\end{subequations}
\end{linenomath*}
In the two equivalent forms for each $I_{k}$, $k=0,1,2$, both $F_{0}$ and
$\psi_{e}$ should be taken as functions of either the electron speed $V$ or
the corresponding kinetic energy, $\mathcal{E}=m_{e}V^{2}/2$, depending on the
integration variable.

It is conventional to express the distribution function and
collision frequencies in terms of the electron kinetic energy, rather than of
the electron speed, so that the integral forms in terms of $\mathcal{E}$ are
more convenient for specific calculations. The form of
equation~(\ref{E_thr_min}) in terms of $I_{k}$ is convenient because it makes
the threshold field totally insensitive to the normalization of $F_{0}$
since, in the relevant fractions, the common coefficients in different $I_{k}$
cancel. This allows one to pick an arbitrary (but common for all $I_{k}$)
normalization of the distribution function, provided $F_{0}$ includes the
entire electron population that consists of the thermal bulk and the
superthermal tail.

If both $\tilde{\psi}=(1+\kappa_{i}^{2}%
)I_{1}/(3I_{0})$ and $\kappa_{i}$ are small (this dual condition is usually
satisfied at altitudes between 100 and 110~km) then equation~(\ref{E_thr_min})
reduces to a simpler relation,
\begin{linenomath*}
\begin{equation}
\frac{(E_{\mathrm{Thr}})_{\min}}{B}=(V_{\mathrm{Thr}})_{\min}\approx\left(
\frac{T_{i}}{m_{i}}+\frac{m_{e}I_{0}I_{2}}{m_{i}I_{1}^{2}}\right)  ^{1/2}.
\label{E_Thr}%
\end{equation}
\end{linenomath*}
In this case, $(E_{\mathrm{Thr}})_{\min}$ becomes insensitive to the
normalization of the function $\psi_{e}$ as well. This allows one to
simultaneously replace in all integrals $I_{k}$ the energy-dependent function
$\psi_{e}$ with merely the $e$-$n$ collision frequency, $\nu_{en}$, so that
for $\psi,\kappa_{i}\ll1$ we have
\begin{linenomath*}
\begin{align}
&  \frac{(E_{\mathrm{Thr}})_{\min}}{B}=(V_{\mathrm{Thr}})_{\min}\approx\left(
\frac{T_{i}+T_{\mathrm{eff}}}{m_{i}}\right)  ^{1/2},\nonumber\\
T_{\mathrm{eff}}  &  =\frac{\left(  \int_{0}^{\infty}F_{0}(\mathcal{E}%
)\sqrt{\mathcal{E}}d\mathcal{E}\right)  \int_{0}^{\infty}F_{0}(\mathcal{E}%
)\left[  d\left(  \mathcal{E}^{5/2}\nu_{en}^{2}(\mathcal{E})\right)
/d\mathcal{E}\right]  d\mathcal{E}}{\left\{  \int_{0}^{\infty}F_{0}%
(\mathcal{E})\left[  d\left(  \mathcal{E}^{3/2}\nu_{en}(\mathcal{E})\right)
/d\mathcal{E}\right]  d\mathcal{E}\right\}  ^{2}}. %
\label{T_eff}
\end{align}
\end{linenomath*}

In the constant-$\nu_{en}$ limit, the effective FB-threshold temperature
reduces to
\begin{linenomath*}
\begin{equation}
\left(  T_{\mathrm{eff}}\right)  _{\nu_{en}=\mathrm{const}}=\frac{10}{9}
\frac{\int_{0}^{\infty}F_{0}(\mathcal{E})\mathcal{E}^{3/2}d\mathcal{E}}
{\int_{0}^{\infty}F_{0}(\mathcal{E})\sqrt{\mathcal{E}}d\mathcal{E}}.%
\label{T_eff_constant_nu_reduces_to}
\end{equation}
\end{linenomath*}
For Maxwellian EDF, $F_{0}(\mathcal{E})\propto\exp(-\mathcal{E}/T_{e0})$, this
further reduces to $(5/3)T_{e0}$, in full accord with
equation~(\ref{gamma_standard}). For the general, non-Max\-wellian EDF, e.g., for
the combined cold bulk electrons and SE,
equation~(\ref{T_eff_constant_nu_reduces_to}) would correspond to merely
including the total electron pressure.

The main result of our FBI linear analysis for general $f_{e}(\vec{V})\approx
F_{0}(V)$ is given by equations~(\ref{omega_r}), (\ref{gamma}), and
(\ref{E_thr_min}); the following relations just represent various
simplifications. It is to be noted, however, that at altitudes closely
approaching the magnetization boundary, $\kappa_{i}=1$, the effect of
ion-thermal instability (ITI) driving becomes tangible \cite{Dimant:Ion04}. The ITI driving modifies both the optimum angles of the instability
onset and the threshold field values. Furthermore, the ITI driving even
extends the unstable range of altitudes by a few kilometers above the
magnetization boundary, where the pure FBI mechanism becomes stabilizing.
Unlike the ETI mechanism, the ITI mechanism destabilizes waves largely in the
same wavelength range as does the FBI\ mechanism, so that the effect of ITI
driving is inseparable from the FBI. We have not included in the present
analysis any thermal effects because that would make our theoretical treatment
much more complicated. This may be a subject of a future work.

\subsection{Specific Calculations for Superthermal Electrons Produced by
Electron Precipitation\label{Specific calculations for superthermal electrons produced by electron precipitation}}

In order to estimate the contribution of the superthermal energy tail formed
by precipitating electrons, we apply the equations derived above to
specific calculations of the FBI threshold. We separate the dominant
omnidirectional part of the total electron velocity distribution,
$f_{e}(\vec{V})\approx F_{0}(\mathcal{E})$, into two distinct components,
\begin{linenomath*}
\begin{equation}
F_{0}(\mathcal{E})\approx F_{\mathrm{TB}}(\mathcal{E})+F_{\mathrm{SE}%
}(\mathcal{E}), \label{F_0(E)}%
\end{equation}
\end{linenomath*}
namely, the undisturbed thermal bulk described by the Maxwellian distribution,
\begin{linenomath*}
\begin{equation}
F_{\mathrm{TB}}(\mathcal{E})=n_{\mathrm{TB}}\left(  \frac{m_{e}}{2\pi
T_{\mathrm{TB}}}\right)  ^{3/2}\exp\left(  -~\frac{\mathcal{E}}{T_{\mathrm{TB}%
}}\right)  , \label{F_TB}
\end{equation}
\end{linenomath*}
and the superthermal EDF, $F_{\mathrm{SE}}(\mathcal{E})$, calculated
numerically using the kinetic code STET, as described above in
sections~\ref{Electron Precipitation in Aurora: General Discussion} and \ref{Results of kinetic simulation}. The
Maxwellian thermal bulk electron distribution, $F_{\mathrm{TB}}$, is fully
determined by the values of the undisturbed temperature, $T_{\mathrm{TB}}$,
and density, $n_{\mathrm{TB}}$. We take these values from ionospheric models,
as described in section~\ref{Results of kinetic simulation}. The superthermal
EDF, $F_{\mathrm{SE}}(\mathcal{E})$, was calculated by STET in the energy
range between 1~eV and 30~keV. In the low-energy range below 1~eV, the main
contributions into all relevant integrals come almost exclusively from
Maxwellian $F_{\mathrm{TB}}(\mathcal{E})$, while the entire energy range above
1~eV is overwhelmingly dominated by $F_{\mathrm{SE}}(\mathcal{E})$. This allows us to
disregard possible inaccuracies of the EDF within the interface energy range
of $\mathcal{E}\sim1$~eV.

For simplicity, we will do our specific calculations for the intermediate
E-region altitudes where both conditions $\psi\ll1$ and $\kappa_{i}\ll1$ hold.
Since $\psi\propto\kappa_{i}^{-1}$, 
there is an overlapping altitude range, roughly between 100 and 110~km, where
both conditions hold concurrently. In this case, the minimum threshold field
is approximately given by equation~(\ref{T_eff}), where normalizations of both
$F_{0}(\mathcal{E})$ and $\nu_{en}(\mathcal{E})$ can be ignored, provided they
are common across all four integrals in the expression for
$T_{\mathrm{eff}}$.

\begin{figure}
\noindent\includegraphics[width=\textwidth]{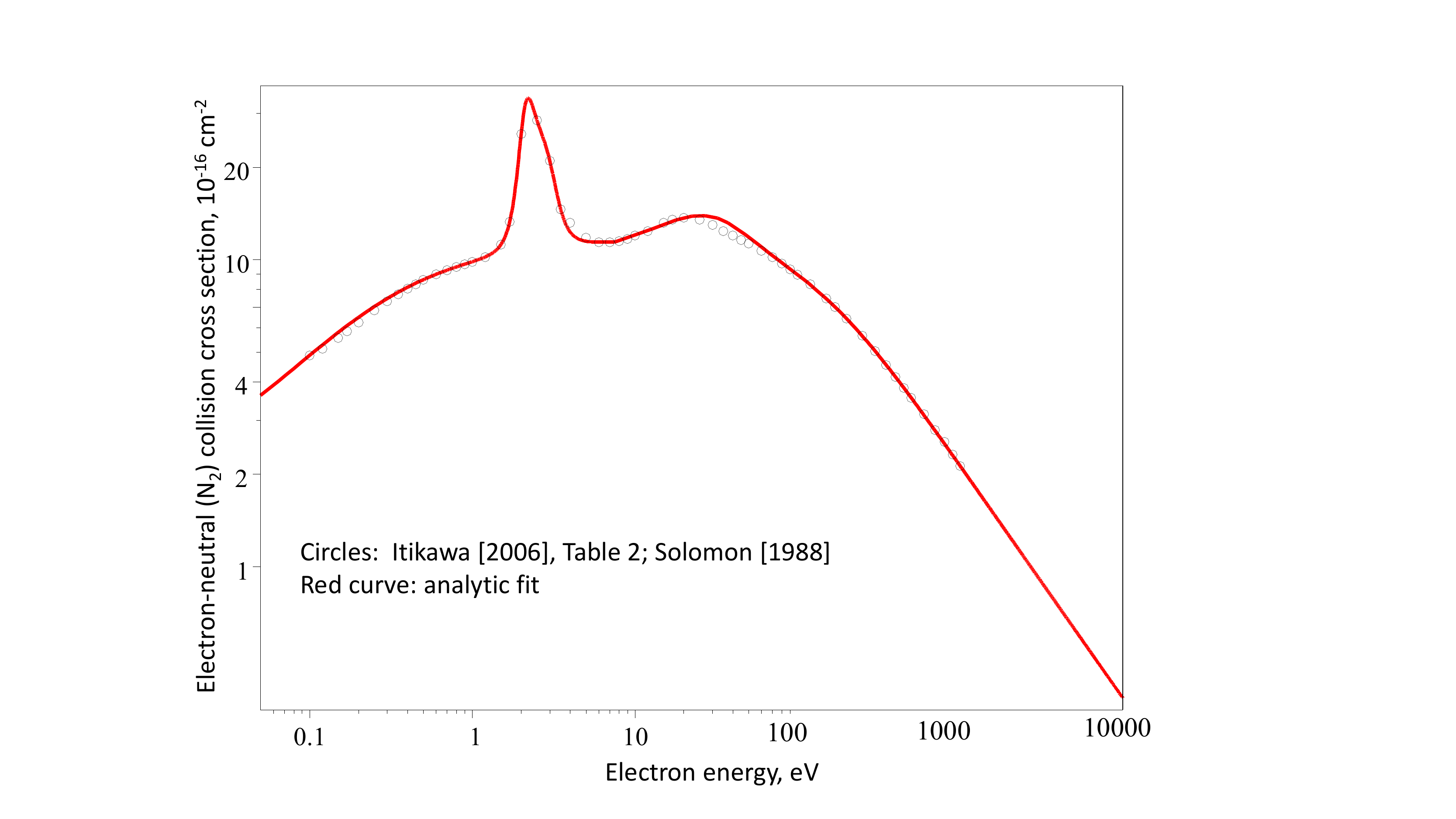}
\caption{The $e$-N$_2$ collision cross-section as a function of the electron energy,
$\mathcal{E}$: \cite[Table 2]{Itikawa:Cross06,Solomon:Auroral88} and analytic fitting,
$\sigma_{en}(\mathcal{E})=10^{-16}\mathrm{cm}^{2}\times\Sigma(\mathcal{E})$,
with $\Sigma(\mathcal{E})$ given by equation~(\ref{Sigma}).}
\label{Fig:Collision_frequency}
\end{figure}
We start by approximating analytically the function $\nu_{en}(\mathcal{E})
=n_{n}\sigma_{en}(\mathcal{E})V(\mathcal{E})$, where $n_{n}$ is the neutral
density and $\sigma_{en}(\mathcal{E})$ is the energy-dependent $e$-$n$
collision momentum transfer cross section. We need to fit $\nu_{en}(\mathcal{E})$
by a continuous analytic function of the electron energy, $\mathcal{E}$, because
the expression for the effective temperature, (\ref{T_eff}), involves the
energy derivative of $\nu_{en}(\mathcal{E})$.

At the altitudes of interest, the
neutral atmosphere consists mostly of the molecular nitrogen and oxygen
($\sim80\%$ of N$_{2}$ and $\sim20\%$ of O$_{2}$), so that N$_{2}$ is more
abundant. Besides, the $e$-$n_{2}$ collisional cross section vastly dominates
over the e-O$_{2}$ collisional cross section \cite<e.g.,>{Solomon:Auroral88}.
This allows us to neglect the e-O$_{2}$ collisions and
approximate the entire neutral population by the nitrogen molecules. In this
approximation, the $e$-$n$ cross section $\sigma_{en}(\mathcal{E})$ becomes a
universal function of the electron energy $\mathcal{E}$, see Fig.~\ref{Fig:Collision_frequency}.
For this paper, we have approximated the data presented in \citeA[Table 2]{Itikawa:Cross06},
with an addition of the top value of $\mathcal{E}=10~$keV
from \citeA{Solomon:Auroral88}. The analytic expression is given by
$\sigma_{en}(\mathcal{E})=10^{-16}\mathrm{cm}^{2}\times\Sigma(\mathcal{E})$,
where the piecewise function $\Sigma(\mathcal{E})$ is expressed in the polynomial-fractional form
$\Sigma(\mathcal{E})=\left.  \sum_{k=0}^{m}\alpha_{k}\mathcal{E}^{k}\right/
\sum_{p=0}^{m}\beta_{p}\mathcal{E}^{p}$ with $\mathcal{E}$ expressed in eV. The details
of this approximation are given in
\ref{Appendix:Analytic approximation of the energy-dependent collision cross-section}.
The universal approximation given by equation~(\ref{Sigma}) can be effectively employed in many kinetic
problems involving collisions of electrons with molecules N$_{2}$.

The effective temperature $T_{\mathrm{eff}}$ is insensitive to the
normalization of the collision frequency $\nu_{en}(\mathcal{E})=
n_{n}\sigma_{en}(\mathcal{E})(2\mathcal{E}/m_{e})^{1/2}$,
so that we can replace the latter with
$\Sigma(\mathcal{E})\sqrt{\mathcal{E}}$, and obtain:
\begin{linenomath*}
\begin{equation}
T_{\mathrm{eff}}=\frac{K_{1}K_{2}}{K_{3}} \label{T_eff_Sigma}%
\end{equation}
\end{linenomath*}
where
\begin{linenomath*}
\begin{align}
K_{1}  &  =\int_{0}^{\infty}F_{0}(\mathcal{E})\sqrt{\mathcal{E}}%
d\mathcal{E},\nonumber\\
K_{2}  &  =\int_{0}^{\infty}F_{0}(\mathcal{E})\ \frac{d\left(  \mathcal{E}%
^{7/2}\Sigma^{2}(\mathcal{E})\right)  }{d\mathcal{E}}\ d\mathcal{E},\label{Integraly}\\
K_{3}  &  =\int_{0}^{\infty}F_{0}(\mathcal{E})\ \frac{d\left(  \mathcal{E}%
^{2}\Sigma(\mathcal{E})\right)  }{d\mathcal{E}}\ d\mathcal{E}.\nonumber%
\end{align}
\end{linenomath*}
and the total distribution function of equation~(\ref{F_0(E)}) can be written
as
\begin{linenomath*}
\begin{equation}
F_{0}(\mathcal{E})=\xi\exp\left(  -\ 31.446\mathcal{E}\right)  +\eta
F_{\mathrm{SE}}(\mathcal{E}),\nonumber
\end{equation}
\end{linenomath*}
with $\xi=3.931\times10^{-7}$, $\eta=1.616\times10^{-19}$,
and $\mathcal{E}$ in eV. The normalization coefficients $\xi$ and $\eta$
provide the distribution function to be measured in $\mathrm{s}^{3}%
\mathrm{m}^{-6}$.

For a specific STET simulation, we pick the Maxwell input with the energy
flux 10~erg~cm$^{-2}$s$^{-1}$, the characteristic energy 30~keV, at the 110~km
of altitude. The simulated electron distribution function in the superthermal
range of energies between 1~eV and 30~keV is reasonably well approximated by a
piecewise expression given by equation~(\ref{pq}).

Using these analytic fits given by equations (\ref{Sigma}) and (\ref{pq}), after all numeric integrations, we obtain
\begin{linenomath*}
\begin{equation}
K_{1}\approx1.976\times10^{-21},\qquad K_{2}\approx2.676\times10^{-19},\qquad
K_{3}\approx3.665\times10^{-21}. \label{K_1-3}%
\end{equation}
\end{linenomath*}
In $K_{1}$ and $K_{3}$, the thermal bulk distribution, $F_{\mathrm{TB}%
}(\mathcal{E})$, vastly dominates the total integrals, whereas in $K_{2}$, due
to the higher power of $\mathcal{E}$ in the integrand, on the contrary, the
superthermal distribution $F_{\mathrm{SE}}(\mathcal{E})$ determines
essentially the entire integral value.

Equation~(\ref{Integraly}) leads to the effective temperature
$T_{\mathrm{eff}}\approx39.4~\mathrm{eV}$ (while the effective electron temperature based on
the total electron pressure, see equation~(\ref{T_eff_constant_nu_reduces_to}),
would yield a much smaller value of $0.11~\mathrm{eV}$). This extremely
high value of $T_{\mathrm{eff}}$ increases the regular FBI threshold
corresponding to $T_{e,i}=300~\mathrm{K}$ (about $0.026~$eV), $E_{\mathrm{Thr}0}
=20[B/(5\times10^{4}nT)]$~mV/m, by a significant factor (almost 30). This results in the enormous
threshold field, $E_{\mathrm{Thr}}\approx\left[  (T_{i}+T_{\mathrm{eff}
})/600~\text{K}\right]  ^{1/2}E_{\mathrm{Thr}0}\approx0.55[B/(5\times10^{4}nT)]~$V/m.
To excite the FBI under these conditions, the convection DC electric
field mapped from magnetosphere down to the E-region altitudes must exceed
this field. At the ionosphere altitudes, such huge convection electric fields,
that would correspond to the $\vec{E}\times\vec{B}$-drift speed as large as
almost 11~km/s, have never been reported. This means that for this level of
precipitation, the strongly elevated FBI threshold can hardly be reached
during extreme geomagnetic storm events, and even during modest ones.

This specific simulation was performed for a relatively strong precipitation
with the mean energy flux $\Phi_{\mathcal{E}}=10~\mathrm{erg}~\mathrm{cm}%
^{-2}~\mathrm{s}^{-1}$. As we discussed in
section~\ref{Results of kinetic simulation}, any superthermal
particle-energy-integrated characteristics will be proportional to
$\Phi_{\mathcal{E}}$. Since the contribution of the superthermal energy tail
to the integrals $K_{1,3}$ is negligible, whereas $K_{2}$ is determined almost
entirely by $F_{\mathrm{TB}}(\mathcal{E})$, the effective FBI\ threshold
\textquotedblleft temperature\textquotedblright, $T_{\mathrm{eff}}$, is in
direct proportion to the energy flux, $T_{\mathrm{eff}}\propto\Phi
_{\mathcal{E}}$. Thus we can generalize the previous result as
$T_{\mathrm{eff}}\approx39.4(\Phi_{\mathcal{E}}/10~\mathrm{erg}~\mathrm{cm}%
^{-2}~\mathrm{s}^{-1})~\mathrm{eV}$.

Furthermore, according to Figure~\ref{Fig:STET}, at a given E-region altitude (e.g., 110~km)
the superthermal EDF does not vary significantly in the broad range of
plasmasheet electron characteristic energies, $\mathcal{E}_{0}$, between 5~keV and 30~keV (at least,
for the Maxwell precipitation input). This allows us to roughly use the
approximation of equation~(\ref{F_0_approx}) to be a `universal' EDF within
the energy domain, say, between 1~eV and a given SE cutoff energy
$\mathcal{E}_{\mathrm{\max}}$ with the zero values outside (in the above
calculation, $\mathcal{E}_{\mathrm{\max}}=20~$keV). This allows us to obtain
an explicit analytic expression for the SE-dominated effective temperature.

\begin{figure}
\noindent\includegraphics[width=\textwidth]{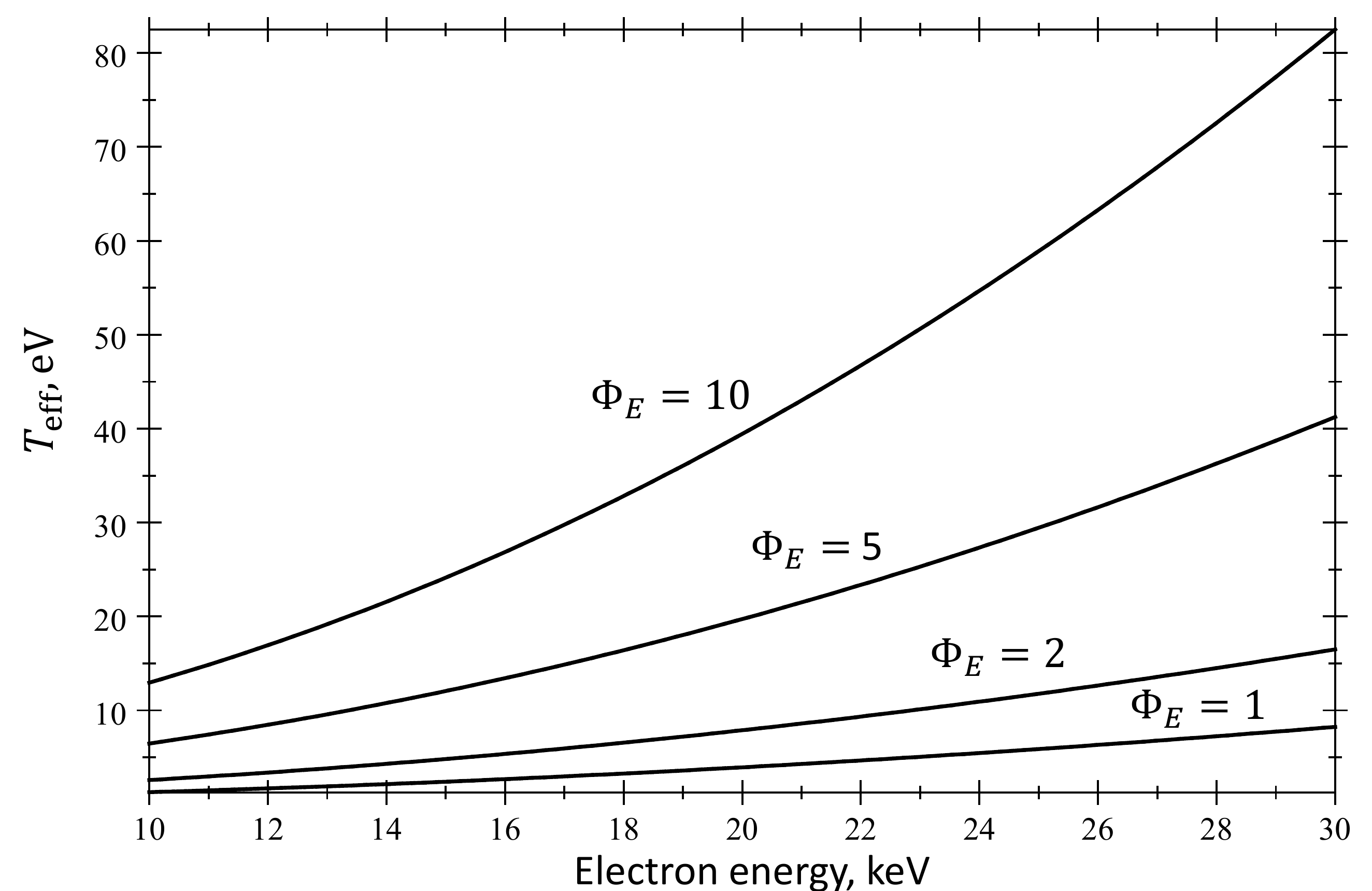}
\caption{Effective temperature (in eV) vs. electron energy (in keV).
The values of the energy flux, $\Phi_{\mathcal{E}}$ (in erg~cm$^{-2}$ s$^{-1}$), are shown near the curves.}
\label{Fig:Teff}
\end{figure}

The idea of this calculation is as follows. Assuming $\mathcal{E}%
_{\mathrm{\max}}$ to be in the energy domain between 1~keV and 30~keV, we can
separate the major integral $K_{2}$ into two parts: a lower-energy part
between 1~eV and 1~keV and the higher-energy remainder. The lower-energy part
can be calculated numerically, which is done above, in equations (), (). This calculation yields a
specific number. For the remaining integral between 1~keV and $\mathcal{E}%
_{\mathrm{\max}}$, we can use the large-energy asymptotics of both functions
$F_{0}(\mathcal{E})$ and $\Sigma(\mathcal{E})$,
\begin{linenomath*}
\begin{align*}
F_{0}(\mathcal{E}) &  =Q(\mathcal{E})\approx\beta\left(  1+\frac
{A}{\mathcal{E}}+\frac{B}{\mathcal{E}^{2}}\right)  ,\\
\Sigma(\mathcal{E}) &  =S(\mathcal{E})\approx m\left(  \frac{1}{\mathcal{E}%
}\right)  ^{\frac{3}{4}}\left(  1+\frac{p}{\mathcal{E}}\right),
\end{align*}
\end{linenomath*}
where
\begin{linenomath*}
\begin{align*}
\beta &  =0.025,\qquad A=634.6,\qquad B=2.721\times10^{6},\\
m &  =373,\qquad p=35.21.
\end{align*}
\end{linenomath*}
For energies above 1~eV these asymptotics are very close to the original
functions. Combining the two parts of the integral, after dropping some small
and inconsequential terms, we obtain for the total SE contribution into
$K_{2}$ a simple algebraic function of $\mathcal{E}_{\mathrm{\max}}$:
\begin{linenomath*}
\begin{align}
&  K_{2}\approx5.622\times10^{-28}\mathcal{E}_{\max}^{2}+7.531\times
10^{-25}\mathcal{E}_{\max}\nonumber\\
&  +~3.085\times10^{-21}\ln\mathcal{E}_{\max}-4.966\times10^{-21}%
.\label{G(E_max)}%
\end{align}
\end{linenomath*}
In the entire energy domain of 1-30~keV, the integrals $K_{1,3}$ are vastly
dominated by the cold bulk-electron energy distribution with the specific
values given by equation~(\ref{K_1-3}), while $K_{2}$, determined almost
entirely by the SE\ energy distribution, is given by equation~(\ref{G(E_max)}).
As a result, we obtain
\begin{linenomath*}
\begin{equation}
T_{\mathrm{eff}}(\mathrm{eV})\approx8.35\times10^{-2}\mathcal{E}_{\max}%
^{2}+0.112\mathcal{E}_{\max}+0.46\ln\mathcal{E}_{\max}%
+2.43,\label{T_eff_general}%
\end{equation}
\end{linenomath*}
where, unlike the above, we express $\mathcal{E}_{\max}$ is in keV. In the
particular case of $\mathcal{E}_{\max}=20$~keV, equation~(\ref{T_eff_general})
reproduces the specific value of $T_{\mathrm{eff}}(\mathrm{eV})\approx
39.4~\mathrm{eV}$ obtained above.
Figure~\ref{Fig:Teff} shows the energy dependence given by equation~(\ref{T_eff_general})
and similar for different values of the energy flux, $\Phi_{\mathcal{E}}$ (the values of
$T_{\mathrm{eff}}$ are proportional to $\Phi_{\mathcal{E}}$).

In this paper, we have restricted our treatment to the FBI, disregarding other
instability drivers, such as the ETI and ITI driving mechanisms. The latter
instabilities may reduce the instability threshold. We may include the other
instability drivers in our future analysis, but one should hardly expect any
drastic changes in the predicted increase of the instability
threshold caused by strong electron precipitation.

In principle, the predicted effect of suppression of the E-region
instabilities by strong electron precipitation is verifiable by observations.
Unfortunately, there are almost no simultaneous collocated observations of
electron precipitation and E-region irregularities caused by the FBI because
such observations and their accurate interpretation represent a certain
challenge [D. Hysell, private communication]. Nevertheless, we are aware of at
least one work where the authors have reported on observations of collocated
optical and radar auroras \cite{Bahcivan:Observations06}. The
data presented there indicate suppression of the FBI inside the auroral arc,
vs. the arc edges where radar aurora still exists. In their conclusion 4,
\citeA{Bahcivan:Observations06} state: \textquotedblleft The
radar aurora was often observed at the discrete arc boundary and suppressed
inside the arc. Radar aurora could sometimes be seen inside an arc at the very
peak of optical intensification.\textquotedblright\ The authors'
interpretation of the observed suppression is that within the arc the driving
electric field might had been dramatically reduced due to the enhanced
conductivity, so that the E-field magnitude might have dropped below the
regular FBI threshold value. While E-field
reduction is a possible mechanism, one must also consider dramatically increased FBI threshold
due to precipitating electrons within the arcs. In order to sustain or
disprove our theory, more future observations with better statistics and more
careful data analysis are needed.

\section{Conclusions\label{Conclusions}}

During events of strong geomagnetic activity, most of the intense
magnetospheric currents close through the high-latitude E-region ionosphere.
At E-region altitudes between 100 and 120~km, strong DC\ electric, mapped down
from the magnetosphere, can drive plasma instabilities, such as the
Farley-Buneman instability (FBI) and others. These instabilities give rise to
anomalous electron heating and enhanced plasma particle transport, affecting
the global ionospheric conductances and, hence, the entire dynamics of the
near-Earth's plasma.

During geomagnetic perturbations, strong electron precipitation also happens,
resulting in Aurora and enhanced ionization. The regions of strong
electric fields that drive E-region instabilities and the auroral regions of
strong electron precipitation may overlap in space, so that the two effects
might interact. This work shows that the intense electron precipitation can modify
significantly the conditions necessary to drive the instability.

We have analyzed theoretically how strong electron precipitation affects
the threshold conditions for the FBI driving. To this end, we performed a
series of kinetic simulations of superthermal electrons, using the
sophisticated kinetic code STET \cite{Khazanov:Non-steady-state93},
using various kinds of the input particle fluxes. These simulations predict distributions of superthermal
electrons in the energy range between 1~eV and 30~keV. While the total number density of the superthermal
electrons (SE) is usually small compared to the total number density of the
electron thermal bulk (TB), the total energy of the entire hot SE
population is often many times that of the entire cold TB population.

This means that the SE total pressure, $P_{\mathrm{SE}}$,
exceeds dramatically the regular pressure of the dominant TB population,
$P_{\mathrm{TB}}=n_{0}T_{e}$. Since the SE particle density is typically small
compared to $n_{0}$ then the dominant SE pressure creates an effective plasma
temperature in proportion to $P_{\mathrm{SE}}$. The elevated electron temperature
increases the particle diffusion and hence the instability threshold, thus
suppressing the instability or at least reducing its efficiency. In order to
quantify the threshold conditions, a naive viewpoint might suggest to just
replace the regular electron temperature with the elevated effective
temperature $P_{\mathrm{SE}}/n_{0}$ in the conventional threshold conditions
for the E-region instabilities.

The actual situation, however, turns out to be more complicated. Even an
oversimplified three-fluid model, in which the TB and SE electron populations
are treated as two different Maxwellian distribution functions, demonstrates
that the FBI threshold field involves the $e$-$n$ collision frequencies whose
values differ dramatically for the two electron populations. The fact that the $e$-$n$
collision frequency varies gradually with energy and cannot be reduced to just two
distinct values means that a quantitative analysis of growth rates requires kinetic theory.

The kinetic theory developed here confirms that
the strongly energy-dependent $e$-$n$ collision frequency plays a crucial
role in the FBI threshold conditions. The physical reason is that the
instability threshold is determined by diffusive losses, where the Pedersen
diffusion (i.e., diffusion along the total electrostatic field and
perpendicular to the magnetic field) plays the principal role in low-frequency
plasma density waves. The Pedersen diffusion coefficient of magnetized
electrons is proportional to the $e$-$n$ collision frequency $\nu_{en}$, so
that for the general non-Maxwellian electron velocity distributions the energy
dependence of $\nu_{en}$ cannot be canceled out. Furthermore, specific
calculations for realistic conditions shows that the energy dependence of
$\nu_{en}$ results in much more severe suppression of the instability compared
to the naive model of the just pressure-dependent threshold: the effective
``temperature'' for the FBI threshold may
exceed that determined by the modified electron pressure alone by more than
order of magnitude.

While there is some observational evidence of the FBI suppression within the
optically active arcs of intense electron precipitation, those observations
cannot be considered as definite proof of our theory because alternative
explanations also exist \cite{Bahcivan:Observations06}.
Nevertheless, we believe that our theoretical treatment is based on solid
physical foundations, so that its major conclusion of possible dramatic
suppression of E-region instabilities by precipitating electrons should be correct.

\appendix
\section{Analytic approximation of the energy-dependent $e$-N$_2$ collision cross-section
\label{Appendix:Analytic approximation of the energy-dependent collision cross-section}}

In this appendix, we approximate analytically the energy-dependent cross-section of electron collisions with N$_2$ molecules.
For $\sigma_{en}(\mathcal{E})$, we use the most up-to-date numerical
model compiled by \citeA{Itikawa:Cross06}, which is in full agreement with the corresponding data published earlier by
\citeA[Fig. A1a]{Solomon:Auroral88}.
\begin{linenomath*}
\begin{equation}
\Sigma(\mathcal{E})=\left\{
\begin{array}
[c]{ccc}%
\frac{a_{0}+a_{1}\mathcal{E}+a_{2}\mathcal{E}^{2}+a_{3}\mathcal{E}^{3}%
+a_{4}\mathcal{E}^{4}+a_{5}\mathcal{E}^{5}}{b_{0}+b_{1}\mathcal{E}%
+b_{2}\mathcal{E}^{2}+b_{3}\mathcal{E}^{3}+b_{4}\mathcal{E}^{4}+b_{5}%
\mathcal{E}^{5}} & \text{if} & \mathcal{E}\leq7.484,\\
&  & \\
\frac{A_{0}+A_{1}\mathcal{E}+A_{2}\mathcal{E}^{2}+373\mathcal{E}^{3}}%
{B_{0}+B_{1}\mathcal{E}+B_{2}\mathcal{E}^{2}+\mathcal{E}^{3.75}} & \text{if} &
\mathcal{E}>7.484.
\end{array}
\right.  , \label{Sigma}%
\end{equation}
\end{linenomath*}
Here the electron energy $\mathcal{E}$ is expressed in eV and the numeric
parameters $a_{k}$, $b_{p}$, $A_{k}$, and $B_{p}$ ($k,p=1,2,3,...$) are given
by
\begin{linenomath*}
\begin{align}
&
\begin{array}
[c]{lll}%
a_{0}=1.0, & a_{1}=29.343\,718\,19, & a_{2}=-48.032\,970\,58,\\
b_{0}=0.570\,827\,0397, & b_{1}=1.773\,322\,602, & b_{2}=-3.690\,608\,199,\\
A_{0}=40281607.78, & A_{1}=-151764.479\,1, & A_{2}=13134.794\,16,
\end{array}
\nonumber\\
&
\begin{array}
[c]{lll}%
a_{3}=28.820\,983\,19, & a_{4}=-7.611\,134\,714, & a_{5}=0.755\,097\,368\,6,\\
b_{3}=2.357\,341\,373, & b_{4}=-0.641\,715\,008\,1, & b_{5}%
=0.06.479\,534\,438,\\
B_{0}=4189550.095, & B_{1}=-117601.051\,2, & B_{2}=3517.072\,043.
\end{array}
\label{abAB_new}%
\end{align}
\end{linenomath*}
Figure~\ref{Fig:Collision_frequency} shows that equation~(\ref{Sigma}) agrees with the published
tabulated data almost perfectly. At the interface energy between the two
pieces, $\mathcal{E}_{c}=7.484~$eV, the function $\Sigma(\mathcal{E}%
_{c})\approx11.41$ is continuous but not smooth; the corresponding derivatives
on both sides of $\mathcal{E}_{c}$ differ by an order of magnitude,
$d\Sigma/d\mathcal{E}|_{\mathcal{E}_{c}-\Delta}\approx0.02$ and $d\Sigma
/d\mathcal{E}|_{\mathcal{E}_{c}+\Delta}\approx0.24$, where $\Delta$ is an
infinitesimal positive number. According to equation~(\ref{T_eff}), the
derivatives of $\Sigma(\mathcal{E})$ are involved in the integrations, but
both values of $d\Sigma/d\mathcal{E}$ around $\mathcal{E}=\mathcal{E}_{c}$ are
so small that the inaccuracy caused by the fitting discontinuity is
inconsequential. It is important that the analytical fit described by
equations~(\ref{Sigma}) and (\ref{abAB_new}) describes adequately all major
details of $\sigma_{en}(\mathcal{E})$, including the well-known N$_{2}$
vibrational excitation peak around 2.5~eV.

\section{Analytic approximation of the SE distribution function
(Maxwell input $\Phi_{\mathcal{E}}=10$~erg~cm$^{-2}$s$^{-1}$, 30~keV, 110~km)
\label{Appendix:Analytic approximation of the SE distribution function}}
\begin{figure}
\noindent\includegraphics[width=\textwidth]{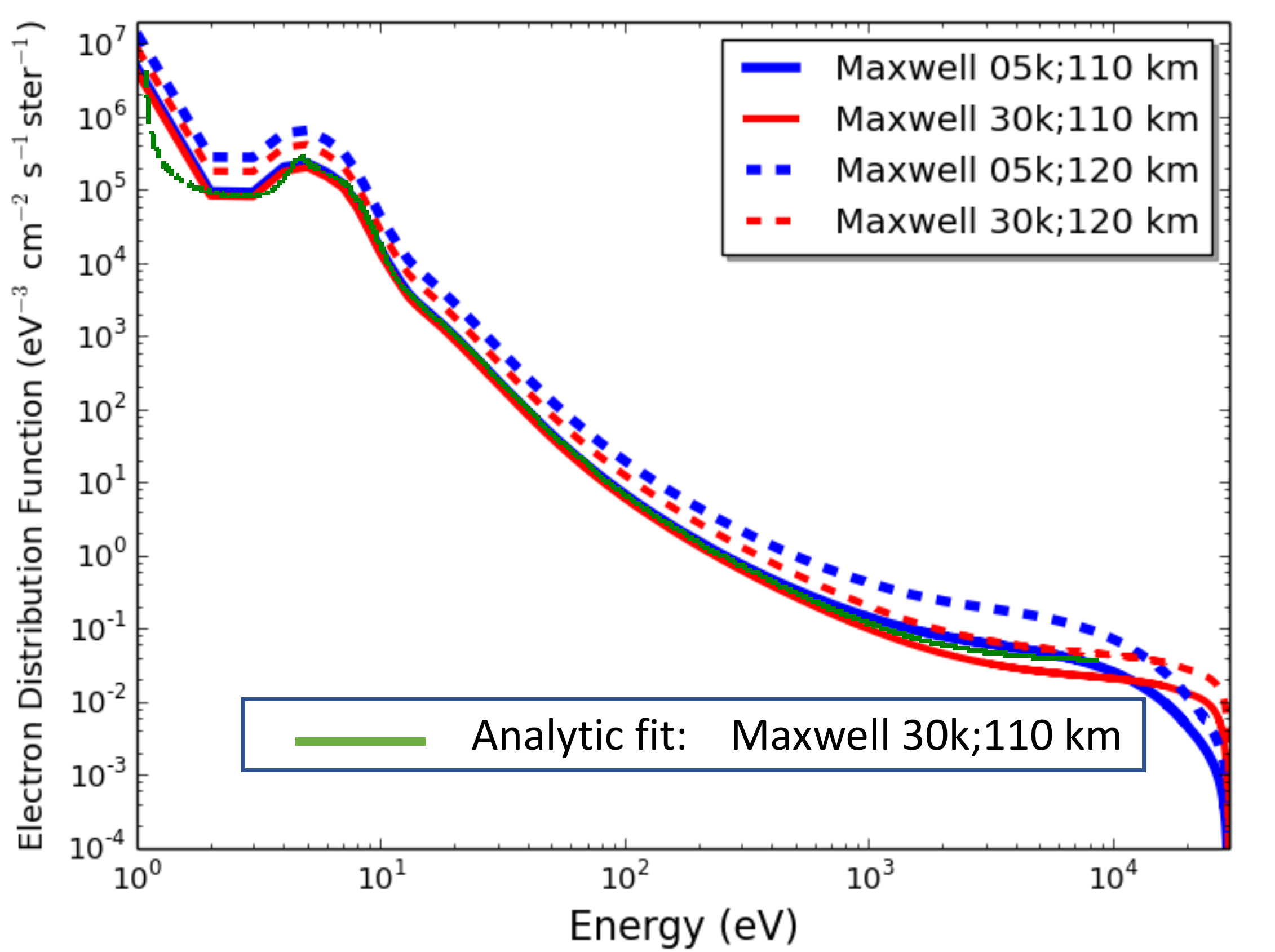}
\caption{Analytic fit of superthermal electron energy distribution function (the green curve); the parameters are shown in the figure.}
\label{Fig:SE_fitting}
\end{figure}

In this appendix, we approximate analytically the STET-simulated distribution function for the Maxwell input with the energy
flux 10~erg~cm$^{-2}$s$^{-1}$, the characteristic energy 30~keV, at the 110~km
of altitude. The simulated electron distribution function in the superthermal
range of energies between 1~eV and 30~keV is reasonably well approximated by a
piecewise expression similar in form to equation~(\ref{Sigma}),
\begin{linenomath*}
\begin{equation}
F_{\mathrm{SE}}(\mathcal{E})\approx\left\{
\begin{array}
[c]{ccc}%
\frac{p_{0}+p_{1}\mathcal{E}+p_{2}\mathcal{E}^{2}+p_{3}\mathcal{E}^{3}%
+p_{4}\mathcal{E}^{4}+p_{5}\mathcal{E}^{5}}{q_{0}+q_{1}\mathcal{E}%
+q_{2}\mathcal{E}^{2}+q_{3}\mathcal{E}^{3}+q_{4}\mathcal{E}^{4}+q_{5}%
\mathcal{E}^{5}} & \text{if} & \mathcal{E}\leq10,\\
&  & \\
\frac{M_{0}+M_{1}\mathcal{E}+M_{2}\mathcal{E}^{2}+M_{3}\mathcal{E}^{3}%
+M_{4}\mathcal{E}^{4}}{N_{0}+N_{1}\mathcal{E}+N_{2}\mathcal{E}^{2}%
+N_{3}\mathcal{E}^{3}+\mathcal{E}^{4}} & \text{if} & \mathcal{E}>10,
\end{array}
\right.  \label{F_0_approx}%
\end{equation}
\end{linenomath*}
where $\mathcal{E}$ is in eV,
\begin{linenomath*}
\begin{equation}%
\begin{array}
[c]{ll}%
p_{0}=1.0, & p_{1}=-7.174\,838\,957,\\
p_{2}=4.278\,098\,241, & p_{3}=-0.936\,524\,197\,8,\\
p_{4}=8.411\,776\,876\times10^{-2}, & p_{5}=-2.740\,658\,004\times10^{-3},\\
q_{0}=1.837\,621\,454\times10^{-4}, & q_{1}=-3.191\,885\,523\times10^{-4},\\
q_{2}=1.713\,056\,278\times10^{-4}, & q_{3}=-4.093\,406\,764\times10^{-5},\\
q_{4}=4.586\,456\,085\times10^{-6}, & q_{5}=-2.000\,3\times10^{-7},
\end{array}
\label{pq}%
\end{equation}
\end{linenomath*}
and
\begin{linenomath*}
\begin{equation}%
\begin{array}
[c]{lll}%
M_{0}=488045957.3, & M_{1}=8068167.375, & M_{2}=73402.046\,4,\\
M_{3}=24.854\,365\,16, & M_{4}=0.025, & N_{0}=-1059681.101,\\
N_{1}=207060.510\,8, & N_{2}=-13361.654\,83, & N_{3}=359.556\,316\,9.
\end{array}
\label{MN}%
\end{equation}
\end{linenomath*}
This analytic fitting is shown in Fig.~\ref{Fig:SE_fitting}. This fitting matches the actual simulated data reasonably well, except the low-energy range of 1-2~eV and, to some degree, above $\mathcal{E}=3$~keV. The former range plays no role, while the latter may introduce some error, but not very significant.

\acknowledgments
Work is supported by NASA LWS Grant \#80NSSC19K0080.


%
%


%
%
%
%
%

\end{document}